\documentclass[12pt]{article}

\usepackage{a4wide,setspace}
\usepackage[pdftex]{graphicx,color}
\usepackage[hang,small,bf]{caption}
\newcommand{\margine}[1]{}
\def\qed{\hspace*{\fill}\rule{2mm}{2mm}}

\newcommand{\junk}[1]{}

\newcommand{\proof}[1]{{\bf Proof.} #1\par}
\newcommand{\mz}[2]{}

\newcommand{\rem}[2]{}

\usepackage{algorithmic}
\usepackage{algorithm}
\usepackage{verbatim}
\usepackage{graphics}
\usepackage{graphicx}
\usepackage{color}
\newtheorem{Le}{Lemma}
\newtheorem{ThS}{Theorem}

\begin{document}
\bibliographystyle{plain}

\title{The Complexity of the Empire Colouring
  Problem\footnote{A preliminary version of this work was presented at
  the 37th International Workshop on Graph-Theoretic Concepts in
  Computer Science (Tepl\'a Monastery, Czech Rep., June 2011).}}

\author{Andrew R. A. McGrae $\quad$ Michele Zito \\
Department of Computer Science, \\ University of Liverpool, \\
Liverpool, L69 3BX, UK \\
e-mail: \texttt {$\{$A.McGrae, M.Zito$\}$@liverpool.ac.uk}}

\maketitle

\begin{abstract}
We investigate the computational complexity of the empire colouring problem (as defined
by Percy Heawood in 1890) for maps containing empires formed by
exactly $r > 1$ countries each. 
We prove that the problem can be solved in polynomial time using
$s$ colours on maps
whose underlying adjacency graph has no
induced subgraph of average degree larger than $s/r$. However, if $s
\geq 3$, the
problem is NP-hard even if the graph is a forest of paths of
arbitrary lengths (for any $r \geq 2$, provided $s < 2r - \sqrt{2r + \frac{1}{4}}+ \frac{3}{2}$). Furthermore
we obtain  a complete characterization of the problem's complexity for
the case when the input graph is a tree, whereas our result for
arbitrary planar graphs fall just short of a similar dichotomy. Specifically, we
prove that the empire colouring problem is NP-hard 
for trees, for any $r \geq 2$, if $3 \leq s \leq 2r-1$ (and
polynomial time solvable otherwise). For  arbitrary planar
graphs we prove NP-hardness if $s<7$ for $r=2$, and $s < 6r-3$, for $r
\geq 3$.
The result for  planar
graphs also proves the NP-hardness of colouring with less than 7 colours
graphs of thickness two and less than $6r-3$ colours
graphs of thickness $r \geq 3$.
\end{abstract}

\section{Introduction}

Let $r$ and $s$ be fixed positive integers.
\margine{Problem definition and relevant notations}
Assume that a partition 
is defined
on the $n$ vertices of a planar graph $G$. In this paper we usually call
the blocks of 
such partition the {\em empires} of $G$ and
we assume that each block contains exactly $r$ vertices.
The graph $G$ along with a partition of this type will be referred
to as an {\em $r$-empire graph}.
The {\em $(s,r)$-colouring} problem
($s$-COL$_r$) asks for a colouring of the vertices of
$G$ that uses at most $s$
colours, never assigns the same colour
to adjacent vertices in different empires
and, conversely, assigns the same colour to all vertices in the 
same empire, disregarding adjacencies.

For $r=1$, the problem coincides with the classical vertex colouring
problem on planar graphs. The generalization for $r \geq 2$ was defined by Heawood 
\cite{heawood90:_map} in the same paper in which he refuted a previous
``proof'' of the famous Four Colour Theorem. It has since been shown  that
 $6r$ colours 
are always sufficient and in some cases necessary to solve 
this problem \cite{jackson84:_solut}. 

In \cite{mcgrae10:_colour_empir_random_trees} (also see
\margine{Previous results}
\cite{mcgrae08:_colour_random_empir_trees}),
we proved that $2r$ colours suffice and are sometimes needed
to colour a collection of empires defined in an arbitrary tree. We also
looked at the proportion of $(s,r)$-colourable trees on $n$ vertices. We
showed that, as $n$ tends to infinity, 
for each $r$ there exists a value $s_r$ such that almost
no tree can be coloured with at most $s_r$ colours and, conversely, for
$s$ sufficiently larger than $s_r$, $s$ colours are sufficient with
(at least) constant positive probability. Later on
\cite{cooper09:_martin_trees_and_empir_chrom}   we improved on this
showing
that, as $n$ tends to infinity, the minimum value $s$ for which 
a random tree is $(s,r)$-colourable is concentrated in a
very short interval with high probability.

Although our investigation considerably expanded the state of
\margine{Main issue: not much is known about the computational
  complexity of $s$-COL$_r$}
knowledge on $s$-COL$_r$, it failed to shed light on its computational 
complexity. Heawood \cite{heawood90:_map} was the first to argue that
there is a simple algorithm that can find 
a $(6r,r)$-colouring in any planar graph $G$ in polynomial time. 
The same process uses at most $2r$ colours if $G$ is a tree.
But what if we only have
$r$ available colours? How difficult is it to decide whether $G$ has an $(r,r)$-colouring?
In this paper we 
\margine{Main results in this paper}
show that $s$-COL$_r$ 
can be solved in polynomial time on planar graphs containing no
induced subgraph of average degree greater than $s/r$. This implies
that,
for instance,  $(2r-1)$-COL$_r$ (resp. $(6r-1)$-COL$_r$) 
can be solved in polynomial time on forests
consisting of paths of length at most $2r-1$ (resp. planar graphs with
components of size at most $12r$). Unfortunately, the outcome of our
investigation seems to indicate that such algorithmic results cannot
be 
extended much further. If $r \geq 2$ and $s \geq 3$, we prove that
$s$-COL$_r$ 
NP-hard on linear forests if $s < 2r - \sqrt{2r + \frac{1}{4}}+ \frac{3}{2}$. Furthermore,
the hardness extends to $s < 6r-3$ (resp. $s< 7$) when $r \geq 3$
(resp. for $r=2$) on arbitrary planar graphs. Finally, for trees, our argument entails a nice dichotomy:
$s$-COL$_r$ is NP-hard 
for any fixed $r \geq 2$, if $s \in \{3, \ldots, 2r-1\}$ and solvable in polynomial time for
any other positive value of $s$.

The
\margine{Quick hints at the reason why our results hold: relationship
  with classical colouring.}
hardness proofs mentioned above
hinge on the fact that the connectivity within
empires has no effect on the graph colourability. 
Essentially, to find an $(s,r)$-colouring in a planar graph $G$, it suffices to be able
to colour with at most $s$ distinct colours (in such a way that no two distinct 
vertices connected by an edge receive the same colour) its {\em reduced graph}
$R_r(G)$. This is a (multi)graph obtained by contracting
each empire to a distinct pseudo-vertex and adding an edge between a pair of pseudo-vertices
$u$ and $v$ for each edge connecting two vertices in the original
graph, one belonging
to the empire represented by $u$, the other one to that represented by
$v$. 
The algorithmic results are based on the use of simple minimum degree greedy
colouring strategies \cite{heawood90:_map} or more refined heuristics
providing algorithmic proofs (see \cite[Theorem 7.9]{G} or
\cite[Exercises 9.12, 9.13]{lovasz93:_combin_probl_exerc}) of the well-known 
Brooks theorem \cite{B41} on such reduced graphs.

The reader at this point may question the reasons 
\margine{Motivations: relationship with classical colouring}
for studying this type of colourings.
Our main interest in the problem comes from its
relationship with other important colouring problems.
Each instance of
$s$-COL$_r$ can be translated to an instance of the classical colouring
problem, but it is not clear to what extent the two problems are
equivalent.
The empire colouring problem is also related to the problem of
colouring graphs of given thickness (a graph has {\em thickness} $t$ \cite{hutchinson93:_color,mutzel98,maekinen09}, if $t$ is the minimum 
integer such that  its edges can
be partitioned into at least $t$ planar graphs). Bipartite graphs can
have high thickness \cite{bhm64}
but only need two colours, and on the other hand a graph of thickness $t$
may have chromatic number as larger as $6t$. Theorem \ref{t2:planar} in this
paper implies that deciding whether a graph of thickness $t \geq 3$ can be
coloured with $s < 6t-3$ colours is NP-hard.

The rest of the paper is organized as follows. 
\margine{Paper plan}
In Section \ref{easy}
we present our positive results concerning sparse planar graphs.  We then
move on (Section \ref{key-red})
to describe a new reduction from the well-known satisfiability
problem to the problem of colouring a particular type of
graph. Hardness results for the colourability of these graphs will be instrumental to our
main results. 
The next Section is devoted to the definition and analysis of a
number of gadgets that will be used in the subsequent reductions. 
Section \ref{paths} deals with the hardness result for forests of
paths. 
The last two sections deal with the hardness results for trees and arbitrary 
planar graphs.

Let $k$ and $s$ be positive integers greater than two.
In what follows $k$-SAT (resp. $s$-COL) denotes the well
known \cite{GJ79,K72} NP-complete problem of checking the satisfiability of a
$k$-CNF boolean formula (resp. deciding whether the vertices of 
a graph $G$ can be coloured using at most $s$ distinct colours in such
a way that no edge of $G$ is monochromatic).  
Also, if $\Pi$ is a decision problem and $\cal
I$ is a particular set of instances for it, then $\Pi({\cal I})$ will
denote the restriction of $\Pi$ to instances belonging to $\cal I$.
If $\Pi_1$ and $\Pi_2$ are decision
problems, then $\Pi_1 \leq_p \Pi_2$ will denote the fact that $\Pi_1$
is polynomial-time reducible to $\Pi_2$. 
Unless otherwise stated we follow \cite{diestel99:_graph_theor}
for all our graph-theoretic notations. 

\section{Algorithms}
\label{easy}

The main outcome of our work is that the empire colouring problem is much
harder than the problem of colouring planar
graphs in the classical sense. However there are cases where things
are easy. 
Let $\sigma$ be an arbitrary positive real number. 
In the following result SPARSE$(\sigma)$ denotes the class of planar
graphs $G$ containing
no induced subgraph of average degree larger than $\sigma$.

\begin{ThS}
\label{positive}
  Let $r$ be an arbitrary positive integer  and $\sigma$ be a positive
  real number such that $r
  \sigma$ is a whole number. The decision
  problem {\rm
    $r \sigma$-COL$_r$(SPARSE$(\sigma)$)} can be solved in polynomial time.
\end{ThS}
\proof{Let $r$ and $\sigma$ be two positive numbers satisfying the
  assumptions above, 
 and assume that
  $G \in$ SPARSE$(\sigma)$, and its vertex set is partitioned into
  empires of size $r$.

If $R_r(G)$ contains a copy of $K_{r\sigma+1}$ then there can be no
$(r\sigma,r)$-colouring of $G$. We now argue that if $R_r(G)$ does not
contain a copy of $K_{r\sigma+1}$ then it is $r \sigma$-colourable (and therefore
$G$ admits an $(r \sigma,r)$-colouring).

Let $S$ be a connected component of $R_r(G)$. In what follows we
denote by $G^S$ the subgraph of $G$ such that $R_r(G^S) \equiv S$. 
 Because all edges of $S$ are edges
in $G^S$, the average degree of this graph satisfies
\[|E(S)|  = |E(G^S)| = \frac{{d}(G^S) \cdot |V(G^S)|}{2}.\]
From this, using the fact that $|V(S)| = |V(G^S)|/r$ and the
definition of SPARSE$(\sigma)$, we have
\[ |E(S)| \leq   \frac{r\sigma}{2} \cdot |V(S)|.\]
This implies that the average degree of $S$ is at most $r \sigma$.
It follows that $S$ is either a regular graph of degree $r \sigma$ or it must
contain at least a vertex of degree less than $r \sigma$. In the former case
$S$ can be coloured with $r \sigma$ colours using, say, the algorithm in
the proof of Brooks' Theorem described in \cite{G}. If $S$ contains
a vertex of degree less than $r \sigma$ we argue that, in fact, the
assumptions about the average degree of all subgraphs of $G$ imply
that any induced subgraph of $S$ is either $r \sigma$-regular or, in turn,
contains a vertex of degree at most $r \sigma-1$. Assume that some
induced subgraph of $S$, $S'$ is not
$r\sigma$-regular and its minimum degree is at least $r \sigma$. This implies that
in particular ${d}(S') \geq r \sigma$. But, by the assumptions
on $G$ the average degree of $S'$ cannot exceed $r \sigma$. Therefore
${d}(S') = r \sigma$ and this implies $S'$ must contain a vertex of
degree less than $r \sigma$.
\qed} 

The result above has a number of interesting consequences.
Let $k$ be a positive integer. Any induced subgraph on $n$ vertices of a forest of paths of
length at most $k$ cannot span more than $k n/(k+1)$ edges. Hence Theorem
\ref{positive} 
implies, for instance, that $\left\lceil\frac{2kr}{k+1}\right\rceil$-COL$_r$ can be
decided
in polynomial time for  forests of paths of
length at most $k$.
Similarly $(6r-1)$-COL$_r$ can be decided in
polynomial time for graphs $G$ formed by arbitrary planar components
of size at most $12r$.

Theorem \ref{positive} also implies that the minimum $s$ for which
$G$ admits an $(s,r)$-colouring can be determined in
polynomial time for any $G \in$ SPARSE$(\sigma)$,
with $r \sigma \leq 3$.

\section{A Useful Reduction}
\label{key-red}

\begin{figure}[htb]
    \centering
    \includegraphics[width=6cm]{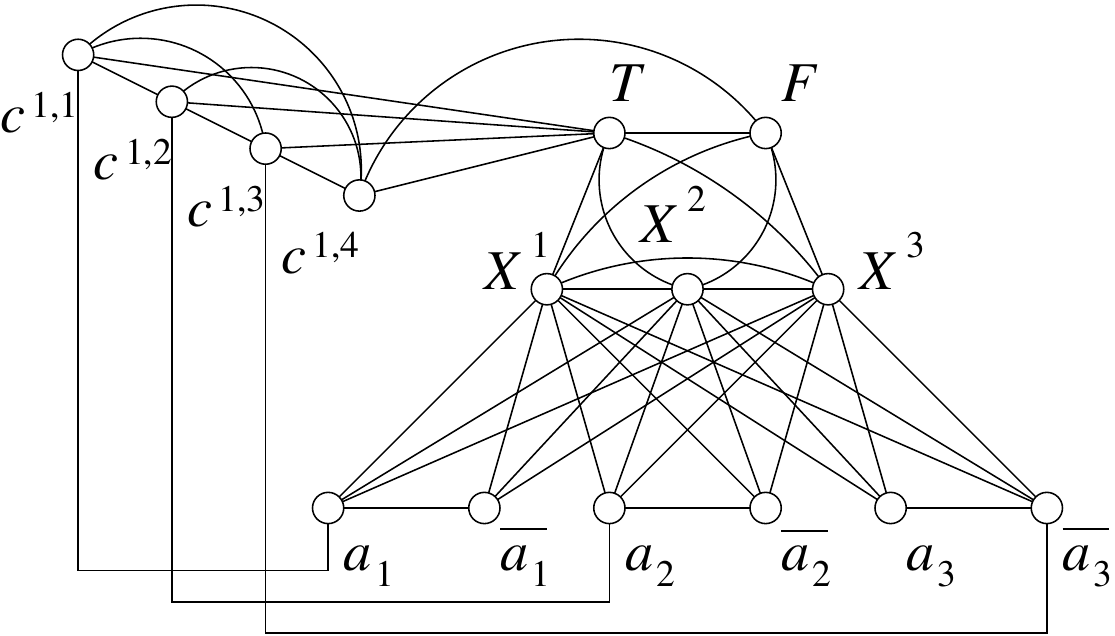}
    \caption{\label{f-g}A small formula graph}    
  \end{figure}

Let $s$ and $k$ be
positive integers with $s > \max(2,k)$. An $(s,k)$-{\em formula graph} is an
undirected graph $\Phi$ such that 
$V(\Phi) = {\cal T} \cup {\cal C} \cup {\cal A}$ where ${\cal T} = \{T,F,X^1, \ldots,
X^{s-2}\}$, ${\cal C}$ contains $m$ groups of vertices 
$\{c^{1,1}, \ldots, c^{1,s-1}\}$, $\{c^{2,1}, \ldots
c^{2,s-1}\}, \ldots, \{c^{m,1}, \ldots, c^{m,s-1}\}$ and 
${\cal A}$ is a set of $2n$ vertices paired
up in some recognizable way. In particular, in what follows we will denote the
elements of ${\cal A}$ by $a_1,\ldots, a_n,\overline{a_1}, \ldots, 
\overline{a_n}$, and we will say that for each $i \in \{1, \ldots,
n\}$, $a_i$ and $\overline{a_i}$ are a {\em pair of complementary vertices}.
Set
${\cal T}$ spans a complete graph; for each pair of complementary
vertices $a$ and $\overline{a}$, 
$\{a,\overline{a},X^j\}$ spans a complete graph for each 
$j \in \{1, \ldots, s-2\}$; for each $i \in \{1, \ldots, m\}$,
$\{T, c^{i,1}, \ldots, c^{i,s-1}\}$ spans a complete graph 
and 
if $j \in \{1, \ldots, k\}$ then
  there is a single edge connecting $c^{i,j}$ to some vertex in ${\cal
    A}$, else if $j \geq k+1$ then $\{c^{i,j}, F\} \in
  E(\Phi)$. Figure \ref{f-g} gives a simple example of a
  $(5,3)$-formula graph.

Let FG$(s,k)$ denote the class of all $(s,k)$-formula graphs. We will now describe
a reduction from $k$-SAT to the problem of colouring using at most $s$
distinct colours the vertices of a given
$(s,k)$-formula graph. The reduction shows the NP-hardness of 
$s$-COL(FG$(s,k)$) for any $k \geq 3$ and $s > k$. 
This in turn will be used repeatedly to prove our hardness
results on $s$-COL$_r$.

\begin{ThS}
\label{fg}
Let $s$ be an integer with $s \geq 3$. Then 
{\rm $k$-SAT $\leq_p s$-COL(FG$(s,k)$)} for any positive integer $k < s$.
\end{ThS}
\proof{Given a $k$-CNF formula $\phi \equiv C_1 \wedge \ldots \wedge
  C_m$ where $C_i$ is the disjunction of $k$ literals ${\sf c}^{i,1},
  \ldots, {\sf c}^{i,k}$
for each $i \in \{1, \ldots, m\}$,
we devise an $(s,k)$-formula graph
  $\Phi$ that admits an
$s$-colouring if and only if $\phi$ is satisfiable. The graph $\Phi$
will consist of one {\sl truth gadget}, one {\sl variable gadget} for
each variable in $\phi$, and 
one {\sl clause gadget} for each clause in $\phi$.

The truth gadget is a complete graph on $s$ vertices labelled $T$,
$F$, and $X^1, \ldots, X^{s-2}$. Note that every vertex in this gadget
must be given a different colour in any $s$-colouring.
Hence w.l.o.g. 
we call these colours ``TRUE", ``FALSE", ``OTHER$^1$", $\ldots$,
``OTHER$^{s-2}$" respectively. 
For each variable $\sf a$ of $\phi$ the variable gadget 
consists of two complementary 
vertices labelled $a$, and $\overline{a}$, connected by an edge and
also adjacent to $X^1, \ldots, X^{s-2}$. 
There are therefore only two ways to colour $a$ and $\overline{a}$:
either $a$ is TRUE and $\overline{a}$ is FALSE or $a$ is FALSE and
$\overline{a}$ is TRUE. Thus the two colourings of $a$ and $\overline{a}$
encode the two truth-assignments of the variable ${\sf a}$.
Each clause
${\sf c}^{i,1} \lor \ldots \lor {\sf c}^{i,k}$ will be represented by $s+k+1$
vertices of $\Phi$. 
Of these, $k$ will
correspond to the clause literals and will be labelled $c^{i,1}, \ldots, c^{i,k}$,
$s-1-k$ will be labelled $c^{i,k+1}, \ldots, c^{i,s-1}$, and the remaining $k + 2$ will be $k$
vertices from variable gadgets and the vertices $T$ and $F$ from the truth gadget.
Vertices $T, c^{i,1}, \ldots, c^{i,s-1}$ form a clique and,
furthermore, for each $j \in \{1, \ldots, k\}$,
 the vertex 
$c^{i,j}$ is connected to the corresponding literal in a variable
gadget. For $k\leq s-2$ vertices $c^{i,j}$, for $j \in \{k+1, \ldots, 
s-1\}$, are adjacent to $F$. Note that, in any colouring of a clause
gadget, vertices $c^{i,j}$, for $j \leq k$, cannot have the same
colour of vertex $T$, and vertices $c^{i,j}$ for $j \geq k$ cannot be
coloured like $F$ either.
The reader can readily verify that $\Phi \in $FG$(s,k)$. The graph in
Figure \ref{f-g} is the $(5,3)$-formula graph corresponding to the
formula
$\phi$ consisting of the single clause ${\sf a}_1 \lor {\sf a}_2 \lor
\overline{{\sf a}_3}$.

If $\phi$ is satisfiable, the elements of ${\cal A}$ in $\Phi$
can be assigned a colour in $\{$TRUE, FALSE$\}$ so that, 
for each $i \in \{1, \ldots, m\}$ at least one
of the $c^{i,j}$ (say for $j=j^*$) is adjacent to some literal coloured TRUE.
This implies that $c^{i,j^*}$ can be coloured FALSE, while all
other $c^{i,j}$
for $j \in \{1, \ldots, s-1\} \setminus \{j^*\}$ can be assigned a
distinct colour in $\{$OTHER$^1,$ OTHER$^2, \ldots,$ OTHER$^{s-2}\}$.
Conversely if there is no way to colour  ${\cal A}$
so that 
for each $i \in \{1, \ldots, m\}$ at least one
of the $c^{i,j}$ is adjacent to some literal coloured TRUE,
then the clause gadget will need $s+1$ colours as the 
$s-1$ vertices $c^{i,j}$ only have $s-2$ colours available (as TRUE
and FALSE are used up by $T$, $F$, and the corresponding literals).
From this we can see that $\Phi$ admits an $s$-colouring 
if and only if there is some way to assign the variables of $\phi$ as
TRUE or FALSE in such a way that every clause contains at least one
TRUE literal.
\qed}

\section{Gadgetry}
\label{G}

Before moving to our hardness results it is convenient to introduce a number of gadgets.

\paragraph{Clique Gadgets.}
Let $r$ and $s$ be positive integers with
$s < 2r$. In what follows the {\em clique gadget}
$B_{r,s}$ 
is an $r$-empire graph satisfying the following properties.
\begin{enumerate}
\item[{\bf B0}] Its graph has $r(s+1)$ vertices partitioned into $s+1$
  empires of size $r$.
\item[{\bf B1}] The graph of $B_{r,s}$ is a forest consisting of $r$ paths.
\item[{\bf B2}] No path in the graph of $B_{r,s}$ contains two vertices from the same empire.
\item[{\bf B3}] The reduced graph of $B_{r,s}$ contains a copy of $K_{s+1}$. Hence
$B_{r,s}$ admits an $(s+1,r)$-colouring and cannot be coloured with
  fewer colours. 
\end{enumerate}

\begin{figure}[htb]
\begin{center}
\includegraphics[angle=0, scale=0.5]{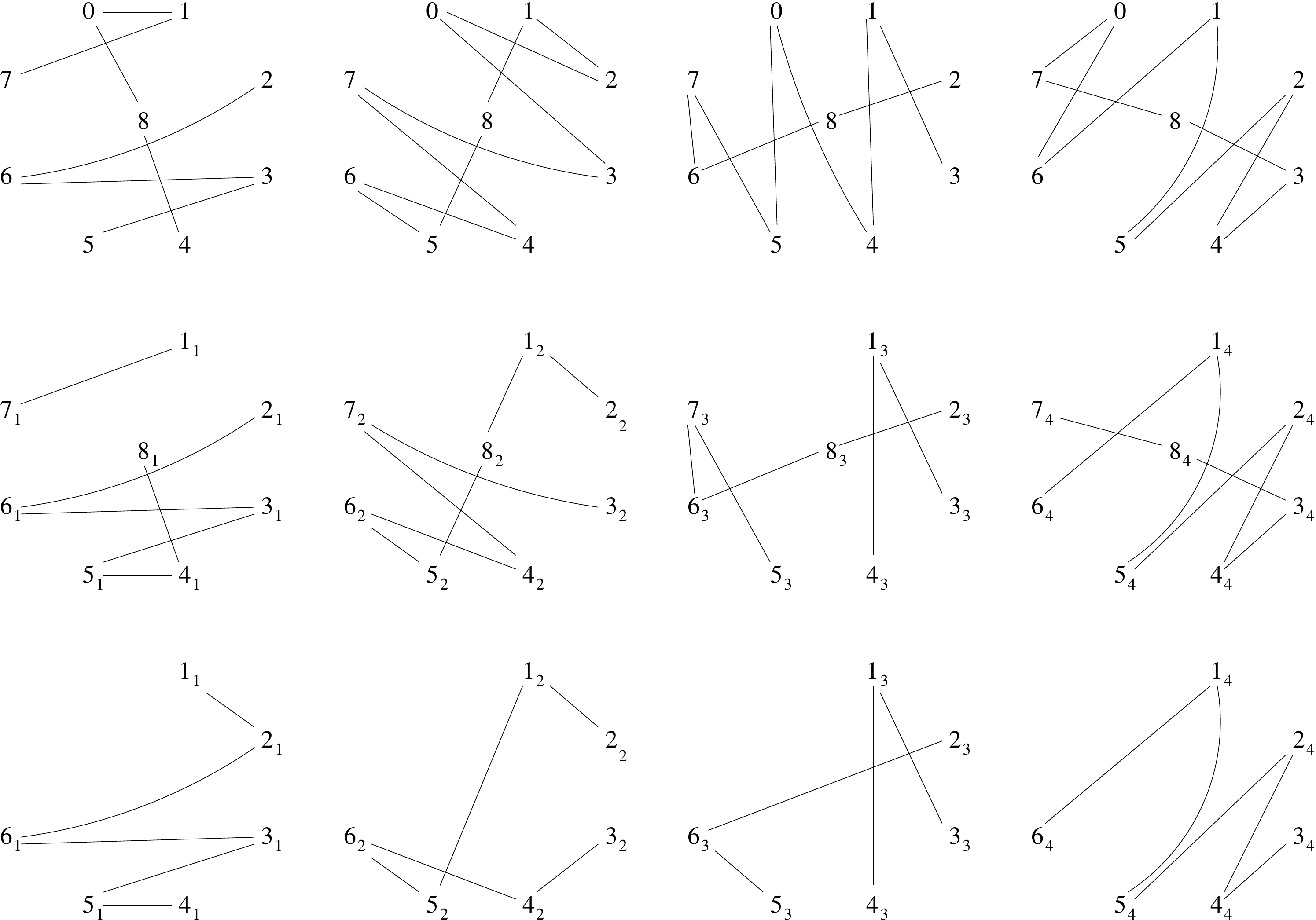}
\end{center}
\vspace{-.3cm}
\caption{\label{FrsPic} Top row: Decomposition of $K_9$ into Hamiltonian cycles. Middle row: $B_{4,7}$. Bottom row: $B_{4,5}$.}
\end{figure}

\begin{ThS}
\label{B}
Let $r$ and $s$ be positive integers with $s < 2r$. Then
there exists an $r$-empire graph $B_{r,s}$ satisfying
properties
{\bf B0}, {\bf B1}, {\bf B2}, {\bf B3}. Furthermore 
$B_{r,s}$ can be constructed in time polynomial in $r$.
\end{ThS}
\proof{For any  positive integer
$r$, the clique $K_{2r+1}$ can be decomposed into $r$ edge-disjoint Hamiltonian
cycles. The result, reported in
\cite[p. 71]{bryant07:_cycle}, is attributed to Walecki (see
\cite{lucas92:_recreat_mathem_vol}). A dummy  $\infty$ is added to the
vertex set of $K_{2r+1}$. The sequence 
\[0, 1, 2r-1,2,2r-2,3, 2r-3, \ldots, r-1,r+1,r,\infty\] 
can be seen as a Hamiltonian cycle of $K_{2r+1}$ after label
``$\infty$'' is identified with vertex $2r$. The remaining cycles are
obtained as cyclic rotations of the first one.

Given one such
decomposition (see top row in Figure \ref{FrsPic}) we define
$B_{r,2r-1}$ (see middle part of Figure \ref{FrsPic}) by copying
cycle $i$ from the decomposition onto vertices $0_i, \ldots, (2r)_i$,
and then taking the induced graph formed by deleting the vertex $0_i$
from the cycle on $0_i, \ldots, (2r)_i$.
Also, if $r>1$, for any $s \in \{1, \ldots, 2r-2\}$ graph of $B_{r,s}$ is obtained 
from that of $B_{r,s+1}$,  by removing all vertices in the empire labelled ${\bf s+2}$ and adding an edge
$\{u,v\}$ whenever $u$ and $v$ are the only two neighbours of
$(s+2)_i$.\qed}

\medskip

\begin{figure}[htb]
\begin{center}
\includegraphics[scale=0.6]{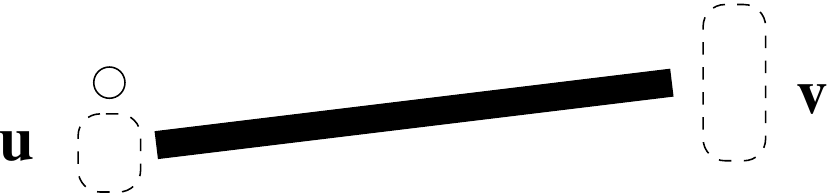}
\caption{\label{ccg} A schematic representation of a $({\bf u},{\bf
    v})$-colour constraining gadget. the diagram shows the isolated
  vertex in empire ${\bf u}$. The two dashed blobs denote,
  respectively, the other vertices in ${\bf u}$ and the vertices in
  ${\bf v}$. The thick black line stands for the part of the gadget
  constraining the colour of ${\bf u}$ and ${\bf v}$: the two empires
  must be given the same colour in any $s$-colouring of $B^-_{r,s}({\bf u}, {\bf v})$.}
\end{center}
\end{figure}

Our results on trees will also need
variants of these gadgets having particular connectivity features. 
Thus if $r>1$ and ${\bf v} \equiv \{v_1,
\ldots, v_r\}$ is some set of $r$ vertices, the 
{\em connected clique gadget rooted at ${\bf v}$},
$B^+_{r,s}({\bf v})$, is formed from $B_{r,s}$, as defined in Theorem
\ref{B}, by adding 
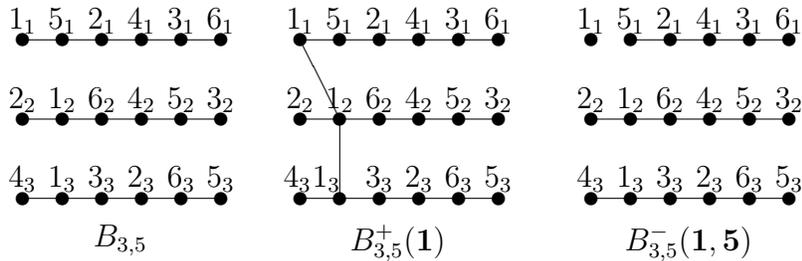
\begin{figure}[h]
\begin{center}
\begin{picture}(300,85)
\put(0,70){\line(1,0){75}}
\put(0,40){\line(1,0){75}}
\put(0,10){\line(1,0){75}}

\put(0,82){\makebox(0,0)[t]{$1_1$}}
\put(15,82){\makebox(0,0)[t]{$5_1$}}
\put(30,82){\makebox(0,0)[t]{$2_1$}}
\put(45,82){\makebox(0,0)[t]{$4_1$}}
\put(60,82){\makebox(0,0)[t]{$3_1$}}
\put(75,82){\makebox(0,0)[t]{$6_1$}}

\put(0,52){\makebox(0,0)[t]{$2_2$}}
\put(15,52){\makebox(0,0)[t]{$1_2$}}
\put(30,52){\makebox(0,0)[t]{$6_2$}}
\put(45,52){\makebox(0,0)[t]{$4_2$}}
\put(60,52){\makebox(0,0)[t]{$5_2$}}
\put(75,52){\makebox(0,0)[t]{$3_2$}}

\put(0,22){\makebox(0,0)[t]{$4_3$}}
\put(15,22){\makebox(0,0)[t]{$1_3$}}
\put(30,22){\makebox(0,0)[t]{$3_3$}}
\put(45,22){\makebox(0,0)[t]{$2_3$}}
\put(60,22){\makebox(0,0)[t]{$6_3$}}
\put(75,22){\makebox(0,0)[t]{$5_3$}}

\put(37,0){\makebox(0,0)[t]{$B_{3,5}$}}

\put(0,70){\circle*{5}}
\put(15,70){\circle*{5}}
\put(30,70){\circle*{5}}
\put(45,70){\circle*{5}}
\put(60,70){\circle*{5}}
\put(75,70){\circle*{5}}

\put(0,40){\circle*{5}}
\put(15,40){\circle*{5}}
\put(30,40){\circle*{5}}
\put(45,40){\circle*{5}}
\put(60,40){\circle*{5}}
\put(75,40){\circle*{5}}

\put(0,10){\circle*{5}}
\put(15,10){\circle*{5}}
\put(30,10){\circle*{5}}
\put(45,10){\circle*{5}}
\put(60,10){\circle*{5}}
\put(75,10){\circle*{5}}

\put(105,70){\line(1,0){75}}
\put(105,40){\line(1,0){75}}
\put(105,10){\line(1,0){75}}

\put(105,70){\line(1,-2){15}}
\put(120,40){\line(0,-1){30}}

\put(105,82){\makebox(0,0)[t]{$1_1$}}
\put(120,82){\makebox(0,0)[t]{$5_1$}}
\put(135,82){\makebox(0,0)[t]{$2_1$}}
\put(150,82){\makebox(0,0)[t]{$4_1$}}
\put(165,82){\makebox(0,0)[t]{$3_1$}}
\put(180,82){\makebox(0,0)[t]{$6_1$}}

\put(105,52){\makebox(0,0)[t]{$2_2$}}
\put(120,52){\makebox(0,0)[t]{$1_2$}}
\put(135,52){\makebox(0,0)[t]{$6_2$}}
\put(150,52){\makebox(0,0)[t]{$4_2$}}
\put(165,52){\makebox(0,0)[t]{$5_2$}}
\put(180,52){\makebox(0,0)[t]{$3_2$}}

\put(104,22){\makebox(0,0)[t]{$4_3$}}
\put(115,22){\makebox(0,0)[t]{$1_3$}}
\put(135,22){\makebox(0,0)[t]{$3_3$}}
\put(150,22){\makebox(0,0)[t]{$2_3$}}
\put(165,22){\makebox(0,0)[t]{$6_3$}}
\put(180,22){\makebox(0,0)[t]{$5_3$}}

\put(142,0){\makebox(0,0)[t]{$B^+_{3,5}({\bf 1})$}}

\put(105,70){\circle*{5}}
\put(120,70){\circle*{5}}
\put(135,70){\circle*{5}}
\put(150,70){\circle*{5}}
\put(165,70){\circle*{5}}
\put(180,70){\circle*{5}}

\put(105,40){\circle*{5}}
\put(120,40){\circle*{5}}
\put(135,40){\circle*{5}}
\put(150,40){\circle*{5}}
\put(165,40){\circle*{5}}
\put(180,40){\circle*{5}}

\put(105,10){\circle*{5}}
\put(120,10){\circle*{5}}
\put(135,10){\circle*{5}}
\put(150,10){\circle*{5}}
\put(165,10){\circle*{5}}
\put(180,10){\circle*{5}}

\put(230,70){\line(1,0){60}}
\put(215,40){\line(1,0){75}}
\put(215,10){\line(1,0){75}}

\put(215,82){\makebox(0,0)[t]{$1_1$}}
\put(230,82){\makebox(0,0)[t]{$5_1$}}
\put(245,82){\makebox(0,0)[t]{$2_1$}}
\put(260,82){\makebox(0,0)[t]{$4_1$}}
\put(275,82){\makebox(0,0)[t]{$3_1$}}
\put(290,82){\makebox(0,0)[t]{$6_1$}}

\put(215,52){\makebox(0,0)[t]{$2_2$}}
\put(230,52){\makebox(0,0)[t]{$1_2$}}
\put(245,52){\makebox(0,0)[t]{$6_2$}}
\put(260,52){\makebox(0,0)[t]{$4_2$}}
\put(275,52){\makebox(0,0)[t]{$5_2$}}
\put(290,52){\makebox(0,0)[t]{$3_2$}}

\put(215,22){\makebox(0,0)[t]{$4_3$}}
\put(230,22){\makebox(0,0)[t]{$1_3$}}
\put(245,22){\makebox(0,0)[t]{$3_3$}}
\put(260,22){\makebox(0,0)[t]{$2_3$}}
\put(275,22){\makebox(0,0)[t]{$6_3$}}
\put(290,22){\makebox(0,0)[t]{$5_3$}}

\put(252,0){\makebox(0,0)[t]{$B^-_{3,5}({\bf 1},{\bf 5})$}}

\put(215,70){\circle*{5}}
\put(230,70){\circle*{5}}
\put(245,70){\circle*{5}}
\put(260,70){\circle*{5}}
\put(275,70){\circle*{5}}
\put(290,70){\circle*{5}}

\put(215,40){\circle*{5}}
\put(230,40){\circle*{5}}
\put(245,40){\circle*{5}}
\put(260,40){\circle*{5}}
\put(275,40){\circle*{5}}
\put(290,40){\circle*{5}}

\put(215,10){\circle*{5}}
\put(230,10){\circle*{5}}
\put(245,10){\circle*{5}}
\put(260,10){\circle*{5}}
\put(275,10){\circle*{5}}
\put(290,10){\circle*{5}}
\end{picture}
\end{center}
\vspace{-.3cm}
\caption{\label{FrsPlusMinus} Examples of clique gadgets.}
\end{figure}
edges $\{v_i,
v_{i+1}\}$ for all $i$ such that $1 \leq i \leq r-1$. Note that the
graph  of such gadget is a tree. Furthermore $B^+_{r,s}({\bf v})$ still satisfies
{\bf B0}, and {\bf B3}.
Finally, if ${\bf u}$ and ${\bf v}$ are two sets of $r$ vertices, the
$({\bf u},{\bf v})$-{\em colour
  constraining gadget} $B^-_{r,s}({\bf u}, {\bf v})$ is an $r$-empire graph
obtained from $B_{r,s}$, without loss of generality,
by removing a single edge connecting the end-point $u_1$ of a path to its
neighbour $v_1$. Thus $u_1$ becomes isolated in the graph of $B^-_{r,s}({\bf u}, {\bf v})$.
The graph
$R_r(B^-_{r,s}({\bf u}, {\bf v}))$  
contains  a copy of $K_{s-1}$ in which every vertex is also
adjacent to the vertices corresponding to ${\bf u}$ 
and ${\bf v}$. 
Thus any $(s,r)$-colouring of $B^-_{r,s}({\bf u}, {\bf v})$ 
must give ${\bf u}$ and ${\bf v}$ the same colour.
Figure \ref{FrsPlusMinus} gives a few examples. 
In the remainder of the paper we will often need to describe
schematically the colour constraining gadgets. Figure \ref{ccg} gives
an example of the graphical notation that will be used.

\paragraph{Connectivity Gadgets.}
\margine{Connectivity gadgets. Initial part, taken from the DM
  submission.}
For positive integers $r$, $s$ and $m$ with $r \geq 2$ and $s \geq 3$,
an $m$-{\em connector},  denoted by 
$A_{r,s,m}$, is an $r$-empire graph satisfying the following
conditions:

\medskip

\mz{2011.8.3}{Do we need the assumption $m \geq r$? If so this needs
  to be stated in some of the forthcoming results.

\medskip

DONE. We don't! If $m \leq r$ a single empty empire will do!

\bigskip

Here's some values of $q$ and $t$ for different values of $r$, $s$,
and $m$. Just in case the referees get nosy!

\begin{center}
\begin{tabular}{llll|ll}
$r$ & range for $s$ & $\sqrt{s-1}$ & $q$ & $m$ & bound on $t$ \\ \hline
2 & $3 \in [3,3.4384472)$ & 1.4142135 & 2 & 1 & $t \geq 0$ \\
 & & & & 2 & $1 + t  \geq 2$ (hence we need $t>0$) \\ 
 & & & & 3 & $1 + t  \geq 3$ (hence we need $t>1$) \\ 
3 & $3 \in [3,5)$ &  1.4142135 & 4 & 1 & $2 + t (12-5) \geq 1$ (any $t>0$
will do) \\ 
& &  & & 2 & $2 + t (12-5) \geq 2$ (any $t>0$
will do) \\ 
& &  & & 3 & $2 + t (12-5) \geq 3$ (any $t>0$
will do) \\ 
& &  & & 4 & $2 + t (12-5) \geq 4$ (any $t>0$
will do) \\ 
& $4 \in [3,5)$ &  1.7320508 & 3 & 1 & $2 + t (12-5) \geq 1$ (any $t>0$
will do) \\ 
\hline
\end{tabular}
\end{center}

}

\begin{figure}
\begin{center}
\includegraphics[scale=0.4]{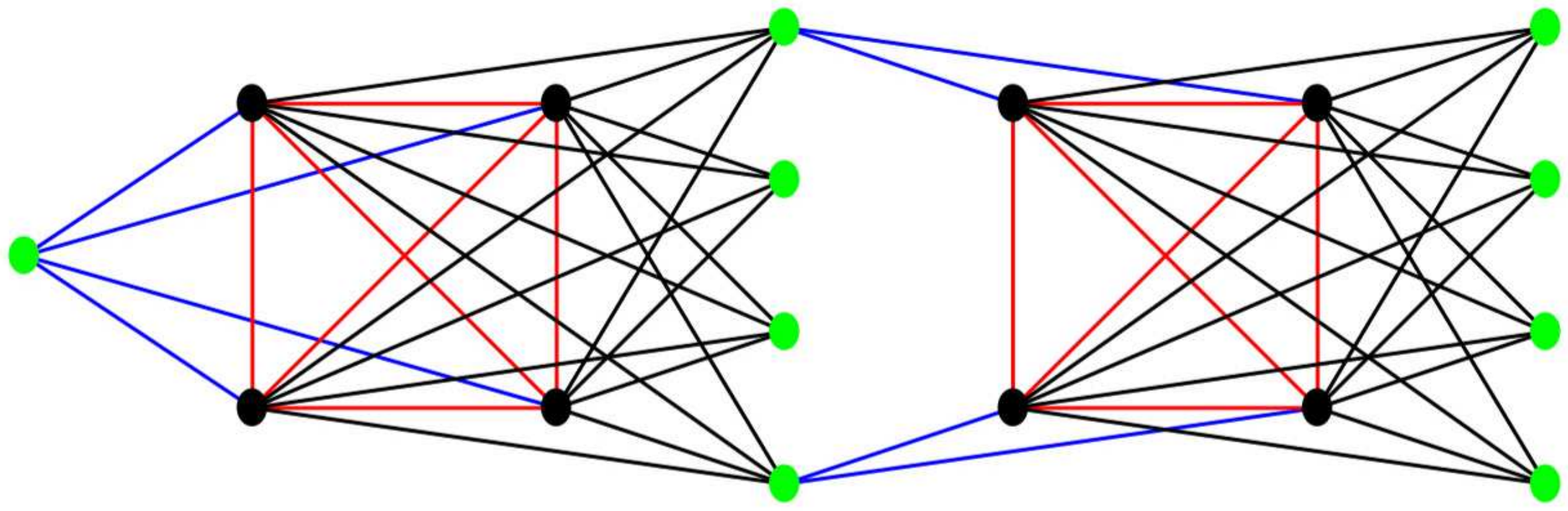}
\caption{\label{EulerGraph} The graph $E_{5,4,2}$, vertices in the monochromatic set are green, edges within the clique are shown in red, and edges connecting a clique to the plug vertex or socket vertices used in its place are blue.}
\end{center}
\end{figure}

\begin{enumerate}
\item[\bf A0] The graph in $A_{r,s,m}$ contains $r \times [1+ (s+q-1)t]$ vertices split into
  empires of size $r$.
\item[\bf A1] The graph in $A_{r,s,m}$ is a linear forest.
\item[\bf A2] There is a set of at least $m$ isolated vertices
in the graph of  $A_{r,s,m}$ 
and such vertices must be given the same colour in any
$(s,r)$-colouring of $A_{r,s,m}$. 
These vertices define the so called {\sl monochromatic set} of the
gadget and will collectively be denoted by $Z$. The elements
of such set will generically denoted by $z$.
\end{enumerate}
Let $q$ and $t$ be arbitrary positive integers. In what follows
\margine{Construction of the connectivity gadgets. Taken from
  Ars6.tex}
$E_{s,q,t}$ is a (non-empire) graph satisfying the following
properties:
\begin{enumerate}
\item[{\bf E0}] $E_{s,q,t}$ contains $(s+q-1)t+1$ vertices.
\item[\bf E1] $E_{s,q,t}$ contains a set of $qt+1$ {\em monochromatic
    vertices}. Each of these  must be given the same colour in any
  proper $s$-colouring of the graph. Among these we identify a {\sl
    plug vertex} which we denote by $u^0$,
and $q$ {\sl socket vertices} denoted by  $u^1, \ldots, u^q$, all of degree 
exactly $s-1$.  The remaining  $q(t-1)$ monochromatic
  vertices are termed {\em internal
    monochromatic vertices}. The remaining  $(s-1)t$ vertices in
  $E_{s,q,t}$ are called {\em colour
  constraining vertices}, and usually denoted by the letter $w$,
appropriately indexed.
\item[\bf E2] The maximum degree of $E_{s,q,t}$ is at most $s+q-1$.
\item[\bf E3] When $s-1$ and $q$ are both even, every vertex in the graph has even degree.
\end{enumerate}
Figure \ref{EulerGraph} shows the graph $E_{5,4,2}$. Graphs
$E_{s,q,t}$ will ``guide'' the constrution of gadgets $A_{r,s,m}({\bf
  v})$ in the sense that for each $r$, $s$, and $m$ there will be values of
$q$ and $t$   such that $E_{s,q,t}$ will be the reduced graph of  $A_{r,s,m}({\bf
  v})$.

\begin{Le}
\label{EGraph}
Let $s$, $q$ and $t$ be positive integers such that $s \geq 3$, and $q
\geq \sqrt{s-1}$. Then there exists a graph $E_{s,q,t}$ satisfying
conditions {\bf E0}, {\bf E1}, {\bf E2}, and {\bf E3}.
\end{Le}
\proof{The graph $E_{s,q,1}$ consists of a {\sl plug vertex} $u^0$, 
$s-1$ {\sl colour constraining vertices} $w^1, \ldots, w^{s-1}$, and 
$q$ {\sl socket vertices} $u^1, \ldots, u^q$. We can see immediately that condition {\bf E0} is satisfied. The edges of $E_{s,q,1}$ are defined as follows: there is a clique on the $s-1$ vertices $w^1, \ldots, w^{s-1}$, also for every $i \in \{0, \ldots, q\}$ and $j \in \{1, \ldots, s-1\}$ there is an edge $\{u^i, w^j\}$.
In any proper $s$-colouring of $E_{s,q,1}$ the vertices $u^0, \ldots, u^q$ must use a colour not used by the $s-1$ vertices in the clique, condition {\bf E1} follows from this. The vertices $w^1, \ldots, w^{s-1}$ all have degree $s+q-1$ while the vertices $u^0, \ldots, u^q$ all have degree $s-1$, conditions {\bf E2} and {\bf E3} follow from this.

For $t>1$, assume that we already have a graph $E_{s,q,t-1}$
satisfying all the required conditions. To create the graph
$E_{s,q,t}$, we add $E_{s,q,1}$, with its plug vertex removed,
to $E_{s,q,t-1}$, and we use the socket vertices of $E_{s,q,t-1}$ to
connect the two graphs. More precisely, the vertices of $E_{s,q,t}$ are
\[V(E_{s,q,t-1}) \cup \left(V(E_{s,q,1}) \backslash \{u^0\}\right).\]
Note that {\bf E0} is satisfied and $E_{s,q,t}$ contains a single plug
vertex and $s-1$ socket vertices. In what follows
$w^1, \ldots, w^{s-1}$ are the $s-1$ colour constraint vertices
belonging to the copy of $E_{s,q,1}$ used to define $E_{s,q,t}$.
The edge set of $E_{s,q,t}$ contains all the edges of $E_{s,q,t-1}$ and $E_{s,q,1} - u^0$ plus $s-1$ additional edges to connect the socket vertices of $E_{s,q,t-1}$ to the colour constraining vertices of $E_{s,q,1}$. 
Each of the colour constraining vertices in $E_{s,q,1}$ is connected to a socket vertex
of  $E_{s,q,t-1}$, in such a way that, after this, the total degree of the socket
vertices is $(q+1)(s-1)$. The assumption $q \geq \sqrt{s-1}$ is needed
at this point, for otherwise the average degree of the socket vertices
would be 
\[\frac{(q+1)(s-1)}{q} = s-1 + (s-1)/q > s-1 + \sqrt{s-1} > s-1+q\] 
where the expression on the right-hand side is the claimed bound on the
maximum degree of $E_{s,q,t}$. Thus, if $q < \sqrt{s-1}$ at least one
of the sockets would have degree larger than $s-1+q$ (thus
contradicting {\bf E2}).

In details, for $s$ odd, we add edges $\{u^{i\ {\rm
    mod}\ q}, w^{2i-1}\}$, and 
$\{u^{i \ {\rm mod}\  q}, w^{2i}\}$ for $i \in \{1, \ldots,
(s-1)/2\}$. Note that we connect an even number of vertices to each socket vertex thus preserving condition {\bf E3}.
For $s$ even, we first add the edge $\{u^i, w^i\}$  for $i \in \{1,
\ldots, \min(s-1,q)\}$. 
If $s-1<q$ some sockets are not used by any of these edges and this completes the
construction of $E_{s,q,t}$. Otherwise 
for $1 \leq i \leq (s-1-q)/2$ we also add edges 
$\{u^{i \ {\rm mod}\  q}, w^{q+2i-1}\}$, $\{u^{i\ {\rm mod}\ q},
w^{q+2i}\}$. Finally, if $q$ is even, we add $\{u^q, w^{s-1}\}$.


As each of the colour constraining vertices of $E_{s,q,1}$ is adjacent
to a socket vertex of $E_{s,q,t-1}$, the clique on these vertices must
use all of the $s-1$ other colours in any proper $s$-colouring. The
socket vertices of $E_{s,q,1}$ must therefore use the one remaining
colour and hence are in the monochromatic set, condition {\bf E1}
follows. \qed}

\begin{ThS}
\label{Arsm}
Let $m$, $r$, and $s$ be positive integers, with $r \geq 2$, and $s$ satisfying
\[3 \leq s < 2r - \sqrt{2r + \frac{1}{4}}+ \frac{3}{2}.\]
Then there exists  a graph
$A_{r,s,m}$ satisfying conditions {\bf A0}, {\bf A1}, and {\bf
  A2}. Furthermore $A_{r,s,m}$  can be constructed in time
polynomial in $r$, $s$ and $m$.
  \end{ThS}
\proof{For $m \leq r$ a single empire of size $r$ with no edges
  satisfies all conditions defining $A_{r,s,m}$. If $m > r$,
we define $A_{r,s,m}$ in such a way that $R_r(A_{r,s,m})$ 
coincides with  
$E_{s,q,t}$, where $q= 2r - (s-1)$ and $t$ is the smallest positive integer such that
\[r-1 + t\left(qr-{(q+1)(s-1) \over 2}\right)  \geq m.\]
Note that the stated bounds on $s$ imply that $q$ satisfies the
conditions of Lemma \ref{EGraph}.

\medskip

\rem{2011.8.5}{For positive values of $s-1$, $2r - (s-1) > \sqrt{s-1}$ is satisfied if $(2r-(s-1))^2
  > s-1$, and the latter is true if $s < 2r -
  \sqrt{2r + \frac{1}{4}}+ \frac{3}{2}$.}

In what follows the empires of $A_{r,s,m}$ will be
denoted by bold type-face letters corresponding to the labels used to
denote the vertices of $E_{s,q,t}$.

\medskip

\rem{2011.7.16}{What is the relationship between $m$ and $q$ and $t$?

DONE}

\bigskip

When $s$ is odd, $q$ is even and hence by condition {\bf E3} every
vertex in $E_{s,q,t}$ has even degree. By a well-known result of Euler
the graph contains an Euler tour, and one such tour can be found in
time polynomial in the size of the graph (see for instance
\cite[Chapter 6]{G}).
 Given one such tour $\cal T$ we can construct the graph
 $A_{r,s,m}$ as follows.
Let $A_{r,s,m}$ be the edgeless graph on $(s+q-1)t+1$ empires
of $r$ vertices, we visit the edges of $\cal T$ 
and add corresponding edges to $A_{r,s,m}$ keeping the invariant that
one of the two end-points of the latest added edge has degree one in
$A_{r,s,,m}$. Without loss of generality
we first add the edge $\{{u^0}_1, {w^1}_1\}$. Then,
assuming we have visited the first $i-1$ edges of $\cal T$ and $v_k$ is the
vertex of degree one incident to the latest added edge, we connect
$v_k$ to an isolated vertex of empire ${\bf u}$, if
$\{v,u\}$ is the next edge we visit in $\cal T$.

The edge set of graph $A_{r,s,m}$ consists of a single long
path, and hence condition {\bf A1} is satisfied. The degree
distribution of  $A_{r,s,m}$ is described in the following
table.

\begin{center}
\begin{tabular}{p{4cm}lll}
vertex set & degree two & degree one & degree zero \\ \hline
${\bf u}^0$ & $\frac{s-1}{2}-1$ & two & $r - \frac{s-1}{2}-1$ \\ \\
{\footnotesize an empire corresponding to a colour constraining vertex} & $r$ &  &  \\ \\
{\footnotesize
one of the $t-1$ groups of $q$ empires corresponding to internal monochromatic vertices} & $(q+1)\frac{s-1}{2}$ & & $qr - (q+1)\frac{s-1}{2}$
\\  \\
${\bf u}^i$ for $i>0$ & $\frac{s-1}{2}$ & & $r
- \frac{s-1}{2}$ \\ \hline
\end{tabular}
\end{center}
Thus $A_{r,s,m}$ has
\[r-1 + t\left(qr-{(q+1)(s-1) \over 2}\right)\]
isolated vertices within the monochromatic set. 
Increasing the value of $t$ will increase this number provided that
\begin{equation}
\label{Eq:odd}
qr > {(q+1)(s-1) \over 2}.
\end{equation}

\bigskip

When $s$ is even, $q = 2r-(s-1)$ is odd. As before, we define
$A_{r,s,m}$ using the graph $E_{s,q,t}$. However this time $E_{s,q,t}$
is not Eulerian.
In particular, all colour constraining vertices have even degree
$s+q-1$. However, 
by the construction used in Lemma \ref{EGraph}, 
in each set of internal monochromatic vertices there
are $\min\{s - 1, q\}$ of even degree. Denote by $u^1, . . . ,
u^{q-s-1}$ the odd degree
vertices in that set. Furthermore, the plug vertex $u^0$, and the final set of $q$ socket vertices are
all of odd degree. To understand the definition of $A_{r,s,m}$ we  define 
a subgraph $H$ of $E_{s,q,t}$. The
edge set of $H$ are defined as follows.
\begin{enumerate}
\item $H$ contains a 
long path $P_0$ starting at $u^0$ and passing through $w^{s-1}$ and $u^q$ of
each set of colour constraining and internal monochromatic vertices.
\item When $s-1 < q$, for each set of internal monochromatic vertices and
  for all $i \in \{1, \ldots, (q-s-1)/2\}$, $H$ contains 
a path $\{u^{2i-1}, w^{i\ {\rm mod} (s-1)}, u^{2i}\}$.
\item Finally, for all $i$ such that $i \leq (q-1)/2$, there is a path
  $\{u^{2i-1}, w^i, u^{2i}\}$, where $u^1, \ldots, u^q$ are the
 socket vertices of $E_{s,q,t}$.
\end{enumerate}

\medskip

\rem{2011.7.27}{Item 2 in the definition of $G_{s,q,t}$: mention that the $w$ empires come from the previous set of colour constraining vertices.

DONE}

\begin{figure}[htb]
\begin{center}
\includegraphics[scale=0.5]{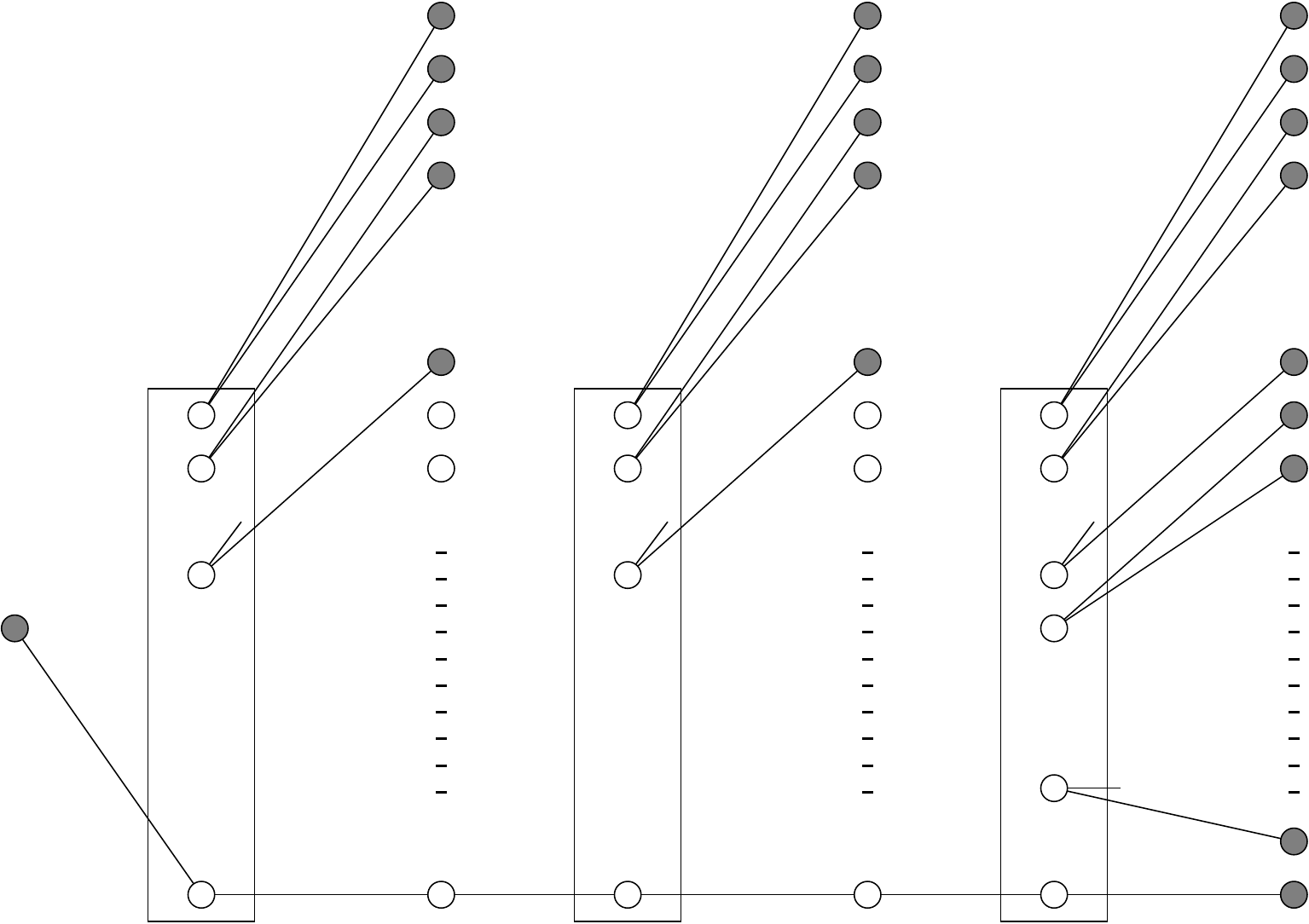}
\end{center}
\caption{\label{G-s-even} The graph $G_{s,q,t}$, when $q>s-1$.}
\end{figure}

Note that $E_{s,q,t} - H$ is Eulerian. We construct $A_{r,s,m}$ in two
stages. We first use the edges of an Euler tour of  $E_{s,q,t} - H$ as
we did in the case $s$ odd. Then we define edges corresponding to the
edges of $H$. This second type of  edges involve different vertices
from those used to deal with the Euler tour.
Finally, if ${u^0}_1$ and ${u^0}_{s/2}$ are the start and the end
point of
the long path in $A_{r,s,m}$ corresponding to $E_{s,q,t} - H$ Euler
tour, and ${u^0}_{s/2+1}$ is the starting point of the path $P_0$ in
$H$, then we can actually attach the edge from ${u^0}_{s/2+1}$ to
${u^0}_{s/2}$. By doing this vertex ${u^0}_{s/2+1}$ becomes isolated,
we lose a vertex of degree one, and gain a vertex of
degree two.

The degree distribution of $A_{r,s,m}$ is given in the
following table.

\medskip

\mz{2011.7.30}{This is Andrew original description:
\begin{itemize}
\item the empire ${\bf u^0}$ has $s/2 -
1$ vertices of degree two, one vertex of degree one and $r-s/2$
isolated vertices. 
\item Each of the $t-1$ sets of $q$ socket empires used
to connect two cliques contain a total of $(q+1)(s-1)/2 -
\max\{(q-s-1)/2, 0\}$ vertices of degree two, $\max\{q-s-1, 0\}$ of
degree one, and $qr-(q+1)(s-1)/2 - \max\{(q-s-1)/2, 0\}$ isolated
vertices. 
\item
The final set of socket vertices contains a total of
$q(s/2-1)$ vertices of degree two, $q$ of degree one, and $q(r-s/2)$
isolated vertices.
\end{itemize}
I think that there is a mistake in the first case (the Euler tour has
to be broken!) ... probably actually the formulas are correct as two
degree one vertices can be identified ... but this should have been mentioned.

Also in the second case the number of vertices of degree two
is
\[(q+1)(s-1)/2 -
\max\{q-s-1, 0\}.\]
}

\begin{center}
{\footnotesize
\begin{tabular}{p{3cm}lll}
vertex set & degree two & degree one & degree zero \\ \hline
${\bf u}^0$ & $\frac{s}{2}-1$ & one & $r - \frac{s}{2}$ \\ \\
an empire corresponding to a colour constraining vertex & $r$ &  &  \\ \\
one of the $t-1$ groups of $q$ empires corresponding to internal
monochromatic vertices &
$(q+1)\frac{s-1}{2} - \max(q-s-1,0)$& $\max(q-s-1,0)$& $qr - (q+1)\frac{s-1}{2}- \max(q-s-1,0)$
\\  \\
${\bf u}^i$ for $i>0$ & $\frac{s}{2}-1$ & one & $r
- \frac{s}{2}$ \\ \hline
\end{tabular}}
\end{center}

In total this gives us
\[(q+1)\left(r-{s \over 2}\right) + (t-1)\left(qr- {(q+1)(s-1) \over 2} - \max(q-s-1, 0)\right)\]
isolated vertices within the monochromatic set. Increasing $t$ will increase this number provided that
\begin{equation}
\label{Eq:even}
qr > {(q+1)(s-1) \over 2} + \max(q-s-1, 0) > 0.
\end{equation}
When $s<r+1$ and hence $\max(q-s-1, 0) = r - s + 1$, the above inequality is always true. We therefore need only consider the case for larger $r$, in this case the bound (\ref{Eq:odd}) on graphs where $s$ is even is the same as the bound when $s$ is odd. The bound can be rewritten as
\[\frac{(s-1)^2}{2} - \left(2r + \frac{1}{2}\right)(s-1) + 2r^2 > 0.\]
This inequality is satisfied for
\[s <2r - \sqrt{2r + \frac{1}{4}}+ \frac{3}{2},\]
and hence for any $m$ and any $s$ and $r$ satisfying the above
inequality, there exists some $A_{r,s,m}$ satisfying conditions {\bf
  A0}, {\bf A1}, and  {\bf
  A2}. \qed}

\medskip

\rem{2011.7.16}{The analysis of the size of the gadget and the number
  fo monochromatic thingies is still missing.

\medskip

WELL, NOT REALLY. $A_{r,s,m}$ has as many empires as the
vertices of $E_{s,q,t}$ ... so that's done. I think that the proof
also gives an estimate on the number of monochromatic vertices.}

\bigskip

Let $r$ be a positive integer. Given 
\margine{The concept of {\em linearization}. Useful to explain how
  connectivity gadgets are used.}
an $r$-empire graph $G$, and an empire $\bf v$ in $G$, the
{\em $r$-degree of ${\bf v}$} is simply the degree of vertex ${\bf v}$
in the reduced graph of $G$ (of course the 1-degree of a vertex in a
graph is just its (ordinary) degree). Let $r'$, $s$, and $m$ be
positive integers as specified at the beginning of this section.
Gadgets $A_{r',s,m}$ will be used in the forthcoming
reductions to replace particular empires with high $r$-degree 
by an array of vertices of degree one or two, chosen among the
monochromatic vertices of the gadget.  Let $m$ be an integer
at least as large as the $r$-degree of ${\bf v}$. The {\em
  linearization of ${\bf v}$ in $G$} 
is the process of replacing ${\bf v}$ in $G$ with a copy of
$A_{r',s,m}$ attaching each edge incident with some element of
$v$ to a distinct element of $Z$ in $A_{r,s,m}$. We
will say that these chosen elements of $Z$ {\em simulate} the
empire ${\bf v}$. Note that, in general, $r'$ may be different from
$r$. Thus repeated linearizations may be used to introduce larger
empires in a given $r$-empire graph or even transform a standard
graph into an $r'$-empire graph, for some fixed $r' > 1$.

\paragraph{Planar Gadgets.}
Let ${\bf u}$ and ${\bf v}$ be given 
set of $r$ vertices and denote by $\delta_{x,y}$ the Kroeneker delta function.
For positive integers $r$, and $s$ with $r \geq 2$ and $s < 6r-3
-2\delta_{r,2}$, it is possible to define a family 
of $r$-empire  graphs $D_{r,s}({\bf u},{\bf v})$ 
satisfying the following properties:
\begin{enumerate}
\item[{\bf D0}] The graph of  $D_{r,s}({\bf u},{\bf v})$  has $r(s+1)$ vertices
  partitioned into $s+1$ empires all of size $r$.
\item[{\bf D1}] The graph of  $D_{r,s}({\bf u},{\bf v})$  contains an isolated vertex $v_1$.
\item[{\bf D2}] No connected component of the graph of $D_{r,s}({\bf u},{\bf v})$ 
contains two vertices from the same empire.
\item[{\bf D3}] The graph $K_{s+1}$ minus the edge $\{{\bf u},{\bf
    v}\}$ is a subgraph of $R_r(D_{r,s}({\bf u},{\bf v}))$.
\end{enumerate} 
$D_{r,s}({\bf u},{\bf v})$ will serve a similar purpose in Theorem
\ref{NPhardPlanar} to that of $B^-_{r,s}({\bf u},{\bf v})$ in
Theorem \ref{NPhard}.

\begin{ThS}
\label{beineke}
Let $r$ and $s$ be positive integers with $r \geq 2$ and $s < 6r-3
-2\delta_{r,2}$. 
Let ${\bf u}$ and ${\bf v}$ be two disjoint sets of $r$
vertices.
 There exists an $r$-empire  graph $D_{r,s}({\bf u},{\bf v})$ 
satisfying conditions {\bf D0}, {\bf D1}, {\bf D2}, and {\bf D3}. Furthermore 
 $D_{r,s}({\bf u},{\bf v})$ can be constructed in time polynomial in $r$.
\end{ThS}
\proof{
 For $r=2$, $s=6$ a suitable graph is shown in Figure \ref{D27}. For $r
 \geq 3$, we can derive $D_{r,6r-4}({\bf u},{\bf v})$ from the proof in
 \cite{beineke65} that the thickness of $K_{6r-3}$ is equal to $r$. In what follow we describe
 Beineke's construction highlighting few points that are important to
 prove properties {\bf D0}, {\bf D1}, {\bf D2}, and {\bf D3}.

\begin{figure}[tb]
\begin{center}
\includegraphics[scale=0.3]{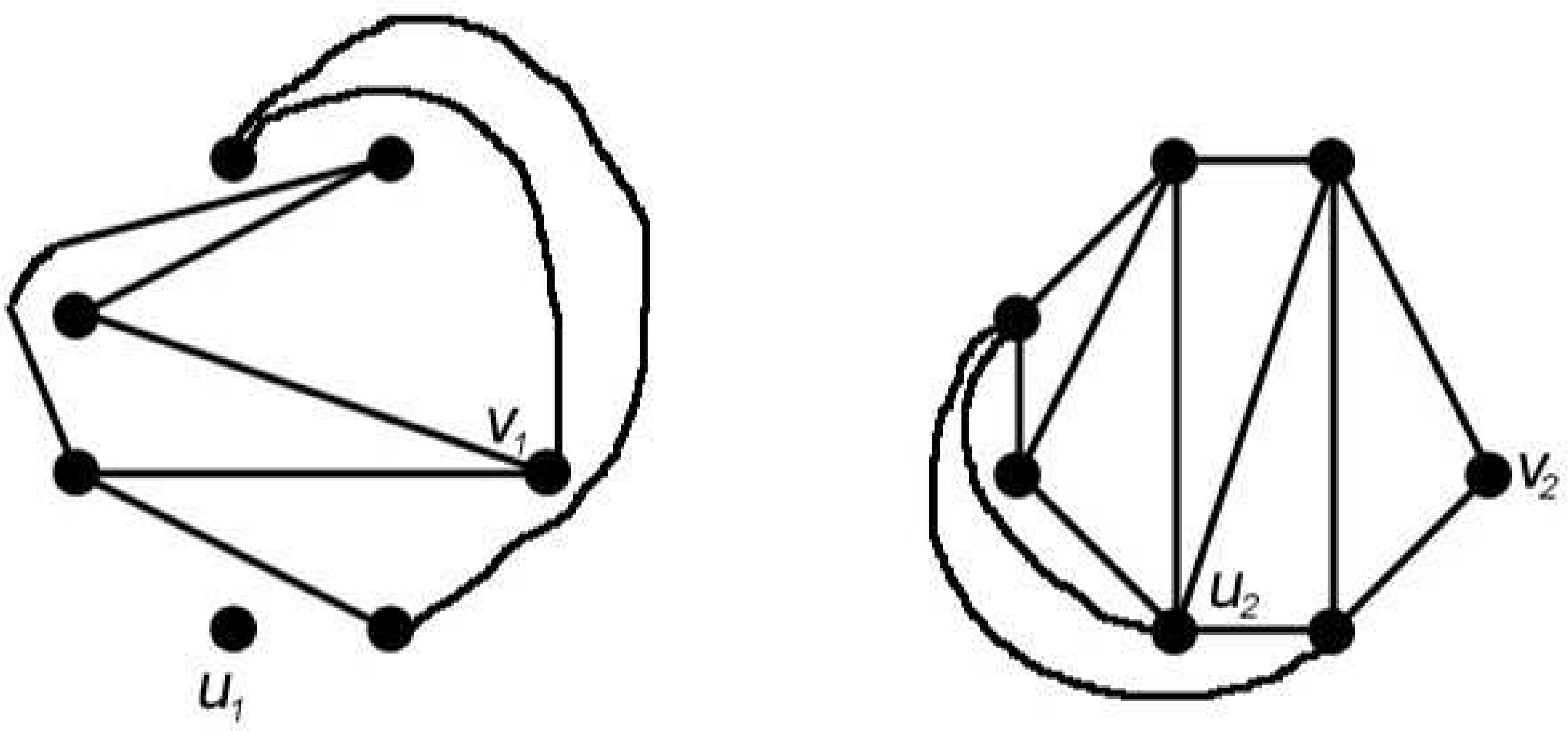}
\end{center}
\caption{\label{D27} The graph $D_{2,6}(u,v)$}
\end{figure}

Beineke's construction starts by showing that there is a graph of
thickness $r-1$ on
 $6(r-1)$ vertices labelled $u(i)$, $v(i)$, $w(i)$, $u'(i)$, $v'(i)$,
 $w'(i)$ for all $i \in \{1, \ldots, r-1\}$ in which there are edges
 connecting every pair of vertices except $\{u(i), u'(i)\}$, $\{v(i),
 v'(i)\}$ and $\{w(i), w'(i)\}$ for each $i \in \{1, \ldots,
 r-1\}$. 

To do this, he defines $D'_{r}$ to be a  graph consisting of $r-1$ connected
 components $G_1, \ldots, G_{r-1}$ (see Figure \ref{Hi}), such that
 for each $i \in \{1 \ldots r-1\}$ $G_i$ consists of $6(r-1)$ vertices
 labelled $u(j)_i$, $v(j)_i$, $w(j)_i$, $u'(j)_i$, $v'(j)_i$,
 $w'(j)_i$ for all $j \in \{1, \ldots, r-1\}$. Vertices $u(i)_i$, $v(i)_i$, $w(i)_i$, $u'(i)_i$, $v'(i)_i$,
 $w'(i)_i$ will be called {\em external}, all others {\em internal} (as
 they are part of a copy of graph $H$). This satisfies
 property {\bf D2}. 

If corresponding vertices in distinct copies of $G_i$ are grouped into
empires of size $r-1$, the reduced graph of 
$D'_{r}$ is a graph meeting Beineke's initial claim. It has $6(r-1)$
 vertices labelled $u(i)$, $v(i)$, $w(i)$, $u'(i)$, $v'(i)$, $w'(i)$
 for all $i \in \{1, \ldots, r-1\}$ 
in which there are edges connecting every pair of vertices except $\{u(i), u'(i)\}$, $\{v(i), v'(i)\}$ and $\{w(i), w'(i)\}$ for each $i \in \{1, \ldots, r-1\}$.

\begin{figure}[tb]
\begin{center}
\includegraphics[scale=0.3]{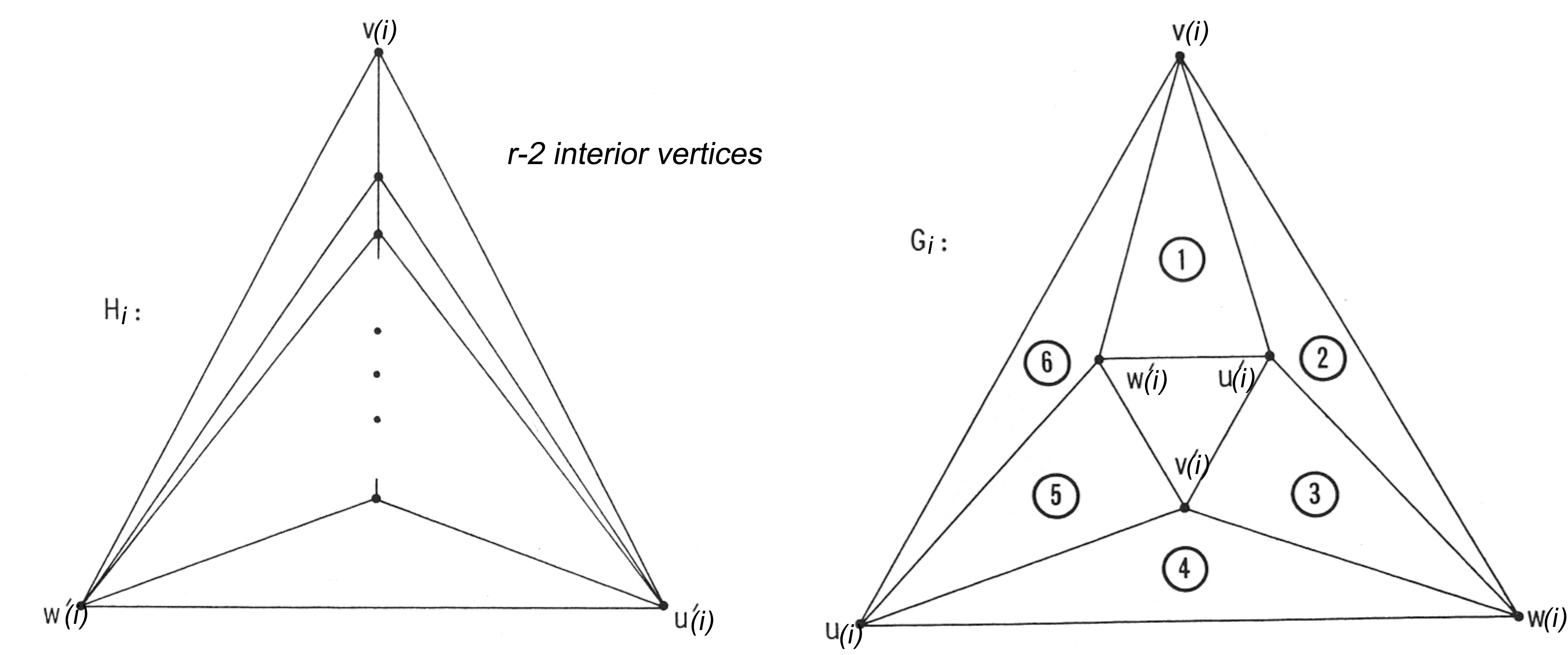}
\end{center}
\caption{\label{Hi} The graphs $H_i$ and $G_i$, the triangles labelled
  $1, \ldots 6$ in $G_i$ contain copies of $H_i$ in which the vertex
  $v_i$ corresponds to $v(i)_i$, $u'(i)_i$, $w(i)_i$, $v'(i)_i$,
  $u(i)_i$, $w'(i)_i$ respectively.  The labelling of the interior
  vertices of the $H_i$ subgraphs is described in \cite{beineke65}.}
\end{figure}

Three more empires $\bf a$, $\bf b$ and $\bf c$, each of
size $r-1$ are added to $D'_r$ 
and connected to it in the following way:
\begin{eqnarray*}
& v\left(\left\lfloor{r-1 \over 2}\right\rfloor\right)_1, v\left(\left\lfloor{r-1 \over 2}\right\rfloor+1\right)_1 & \textrm{ are adjacent to } a_1\\
& u(i)_i, u'(i+1)_i, v(i)_i & \textrm{ are adjacent to } a_i (i>1), \\ \\
& v(1)_1, v(2)_1, u'(1)_1 & \textrm{ are adjacent to } b_1\\
& u(1)_{\lceil{r-1 \over 2}\rceil + 1}, u'(2)_{\lceil{r-1 \over 2}\rceil + 1} & \textrm{ are adjacent to } b_{\lceil{r-1 \over 2}\rceil + 1}\\
& v'(i)_i, v(i+1)_i, u(i)_i & \textrm{ are adjacent to } b_i (i \in \{1, \ldots, \lceil{r-1 \over 2}\rceil\})\\
& v(i)_i, v'(i+1)_i, u'(i)_i & \textrm{ are adjacent to } b_i (i \in
\{\lceil{r-1 \over 2}\rceil+2, \ldots, r-1\}), \\ \\ 
& w'(2)_1 & \textrm{ is adjacent to } c_1\\
& w(i)_i, w'(i+1)_i & \textrm{ are adjacent to } c_i (i>1).
\end{eqnarray*}
As each vertex from empires $\bf a$, $\bf b$ and $\bf c$ was added to a single
component of $D'_{r}$, property {\bf D2} is still satisfied. Let
$G_r$ be the complement of $R_{r-1}(D'_r + \{{\bf a},{\bf b},{\bf
  c}\})$. It is not difficult to see that 
$G_r$ is planar. 
Therefore by adding the vertices $u(j)_r$, $v(j)_r$, $w(j)_r$,
$u'(j)_r$, $v'(j)_r$, $w'(j)_r$ for all $j \in \{1, \ldots, r-1\}$
with the same edge set as $G_r$ we have a graph consisting of $r$
planar components 
(that's $G_r$ along with the components of the augmented graph $D'_r + \{{\bf a},{\bf b},{\bf
  c}\}$)
that reduces to $K_{6r-3}$.

$G_1$ contains a vertex $c_1$ of degree one which is adjacent to
$w'(2)_1$. We can now form the graph $D_{r,6r-4}({\bf u}, {\bf v})$ from $G_r$ along with the components of the augmented graph $D'_r + \{{\bf a},{\bf b},{\bf
  c}\}$ by
renaming empires $c$ and $w'(2)$ as $\bf v$ and $\bf u$ respectively and
removing all edges between $\bf u$ and $\bf v$. $D_{r,6r-4}({\bf u},
{\bf v})$ satisfies
property {\bf D0},  {\bf D1} (as the only edge incident to $\bf v$
has been deleted), {\bf D2}
and {\bf D3} as the graph reduces to $K_{6r-3}$ minus the edge
$\{u,v\}$. For $s < 6r-4$, note that the induced graph formed by
removing any empire other than $\bf u$ or $\bf v$ 
from $D_{r,s+1}({\bf u}, {\bf v})$ is an example of $D_{r,s}({\bf
  u},{\bf v})$. 
As the size of the graph $D_{r,s}({\bf u}, {\bf v})$ depends only on $r$ and $s$, the graph can be constructed in polynomial time.
\qed}

\section{Linear Forests}
\label{paths}

In Section \ref{easy} we showed (amongst other things) that there are specific values for 
$s$ such that $s$-COL$_r$  becomes easy if the
input graph is a collection short paths. Here we argue that  if the
paths are allowed to have arbitrary length (let LFOREST denote the set
of all forests of this form) then the problem becomes NP-hard.
We will prove the following result.

\begin{ThS}
\label{Paths}
Let $r$ and $s$ be positive integers with $r \geq 2$ and $3 \leq s <  2r - \sqrt{2r + \frac{1}{4}}+ \frac{3}{2}$. Then
the 
{\rm $s$-COL$_r$(LFOREST)}
problem is {\rm NP}-hard.
\end{ThS}

Note that it follows from results in
\cite{mcgrae08:_colour_random_empir_trees} 
that any $r$-empire graph
defined on a linear forest can be coloured in polynomial time using
$2r$ colours. Thus Theorem \ref{Paths} is, at least for large values
of $r$, close to best possible, in the sense that the largest values
of $s$ for which it holds are $2r-1+o(r)$.

The proof is split into two parts. The argument for $s=3$ is based on a
direct construction which is reminiscent of a well-known hardness proof
for $3$-COL \cite[p.1103]{cormen09:_introd_algor}. For $s>3$, 
the hardness of $s$-COL$_r$(LFOREST) will then follow
from that of $s$-COL(FG$(s,s-1)$).

We start
\margine{Case $s=3$ same proof as DM. Changed it a bit (2011.7.29).}
from the case $s=3$.
\begin{ThS}
\label{r2s3}
Let $r$ be an integer with $r \geq 2$. Then {\rm 3-SAT $\leq_p$ 3-COL$_r$(LFOREST)}. 
\end{ThS}
\proof{
The proof construction is reminiscent of that used to show that
  3-COL is NP-hard \cite[p.1103]{cormen09:_introd_algor}. 

Given an instance $\phi$ of 3-SAT we can produce a linear
  forest 
$P(\phi)$ and a partition of $V(P(\phi))$ into empires of size
  $r$ such that $P(\phi)$ admits a 
$(3,r)$-colouring if and only if $\phi$ is satisfiable. 
$P(\phi)$ consists of one {\sl truth gadget}, one {\sl variable gadget} for each variable used in $\phi$, and one {\sl clause gadget} for each clause in $\phi$.

\begin{figure}[htb]
\begin{center}
\includegraphics[scale=0.4]{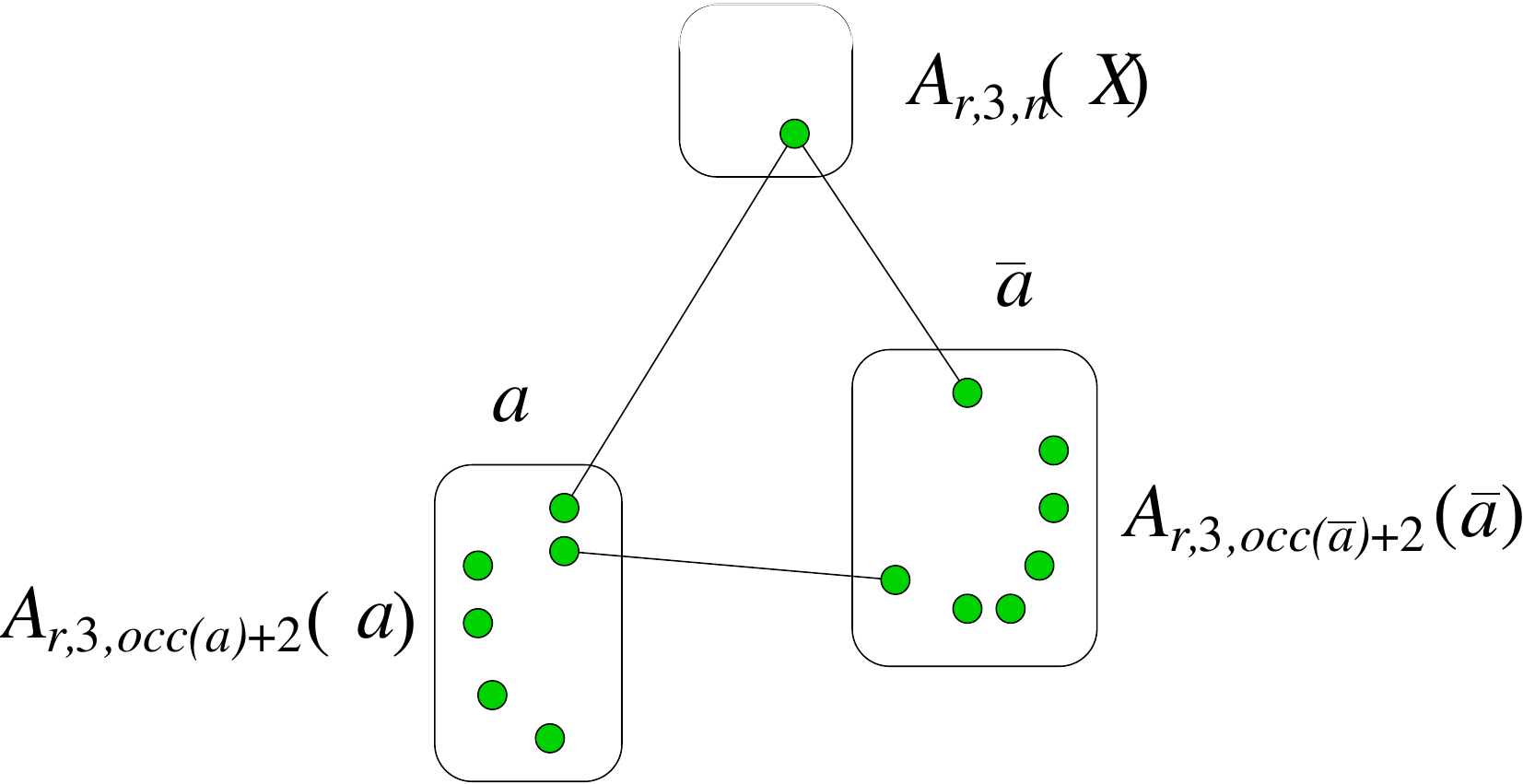}
\end{center}
\vspace{-.3cm}
\caption{\label{varS3} The shape of a variable gadget for
  $s=3$.}
\end{figure}
To define the truth gadget, we start by adding $r-2$
distinct isolated vertices to each empire in $B_{2,2}$.  
The empires in
the resulting graph (which we denote by $B_{2,2}^{+r}$) will be labelled 
${\bf T}$, ${\bf F}$ and ${\bf X}$.
Then, if $\phi$ uses $n$ different variables and $m$ clauses, we
linearize {\bf T} and {\bf X} in $B_{2,2}^{+r}$, using one copy of 
$A_{r,3,\deg({\bf T})+2m}$, and one copy of $A_{r,3,\deg({\bf
    X})+n}$
(here $\deg({\bf v})$ is the degree of empire ${\bf v}$ in
$B_{2,2}^{+r}$), respectively. We denote such gadgets by $A({\bf T})$
and
$A({\bf X})$ respectively. This
completes the definition of the truth gadget. Since ${\bf T}$, ${\bf F}$ and ${\bf X}$ 
are all adjacent (in $B_{2,2}^{+r}$) and the linearization preserves
colour constraints (because of property {\bf A2}), the vertices of the
truth gadget simulating the three empires of $B_{2,2}^{+r}$
must have different colours in any 3-colouring of the truth
gadget. Without loss of generality we call TRUE, FALSE and OTHER
respectively such colours.

For each variable ${\sf a}$ in $\phi$, $P(\phi)$ contains a variable
gadget. Let occ$( \cdot )$ be a function taking as input a literal of $\phi$ and 
returning the number of occurrences of
its argument in the given formula.
The variable gadget for {\sf a} is defined as the graph formed by the
two connectivity gadgets $A_{r,3,{\rm occ}({\sf a})+2}$ and 
$A_{r,3,{\rm occ}(\overline{\sf a})+2}$, along with
a single monochromatic vertex $z$ in $A({\bf X})$ (a distinct
monochromatic vertex is used for each variable of $\phi$). The edges in
the variable gadgets will be those of $A_{r,3,{\rm occ}({\sf a})+2}$ and 
$A_{r,3,{\rm occ}(\overline{\sf a})+2}$ plus three
further edges: $\{z,z_{\bf a}\}$, $\{z,z_{\bf\overline{a}}\}$, and 
 $\{z'_{\bf a},z'_{\bf\overline{a}}\}$. Here $z_{\bf a}$ and  $z'_{\bf
   a}$
(resp. $z_{\bf\overline{a}}$ and $z'_{\bf\overline{a}}$) are 
monochromatic vertices in $A_{r,3,{\rm occ}({\sf a})+2}$ and 
$A_{r,3,{\rm occ}(\overline{\sf a})+2}$.
Figure \ref{varS3} gives a schematic view of 
the truth gadget for an arbitrary variable ${\sf a}$.
Since ${\bf X}$ has colour OTHER, there are only two possible
colourings for the vertices corresponding to 
${\bf a}$ and ${\bf \overline{a}}$ --- either all vertices for ${\bf
  a}$ are coloured  TRUE and those for
${\bf \overline{a}}$ are coloured 
FALSE, or the vertices for ${\bf a}$ are coloured FALSE and those for ${\bf \overline{a}}$ TRUE. 

Finally, for each clause in $\phi$, $P(\phi)$ contains a gadget
like the one depicted in Figure \ref{PathClauseGadget}. This is
connected to the rest of the graph via four connectivity gadgets. More
specifically,
the two vertices labelled $T_1$ and $T_2$ (in the Figure) are two monochromatic vertices in 
$A({\bf T})$ (a distinct pair of such  monochromatic vertices
for each case clause gadget). Also, vertices labelled $a$, $b$ and $d$
in the Figure belong to the monochromatic set of three connectivity 
gadgets of the form $A_{r,3,{\rm occ}({\sf
    \ell})+2}$ where $\ell$ is a literal (${\bf \ell} = {\sf
  a}$, ${\sf b}$, and ${\sf d}$ in the given example).
Since the vertices of $A(T)$ corresponding to ${\bf T}$ will always be coloured TRUE, it
can be shown that each clause gadget admits a proper
$(3,r)$-colouring if and only if at least one of the empires 
corresponding to a literal in the clause is
coloured TRUE. 

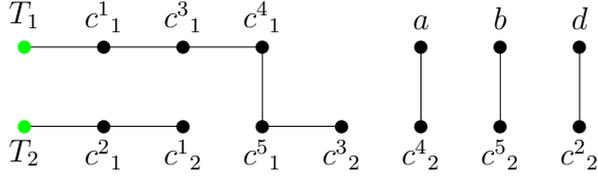
\begin{figure}[t]
\begin{center}
\begin{picture}(230,45)

\put(10,35){\line(1,0){30}}
\put(10,5){\line(1,0){30}}
\put(40,35){\line(1,0){30}}
\put(70,35){\line(1,0){30}}
\put(40,5){\line(1,0){30}}
\put(100,5){\line(0,1){30}}
\put(100,5){\line(1,0){30}}
\put(160,5){\line(0,1){30}}
\put(190,5){\line(0,1){30}}
\put(220,5){\line(0,1){30}}

\put(10,52){\makebox(0,0)[t]{$T_1$}}
\put(10,0){\makebox(0,0)[t]{$T_2$}}
\put(40,0){\makebox(0,0)[t]{${c^2}_1$}}
\put(40,52){\makebox(0,0)[t]{${c^1}_1$}}
\put(70,0){\makebox(0,0)[t]{${c^1}_2$}}
\put(70,52){\makebox(0,0)[t]{${c^3}_1$}}
\put(100,0){\makebox(0,0)[t]{${c^5}_1$}}
\put(100,52){\makebox(0,0)[t]{${c^4}_1$}}
\put(130,0){\makebox(0,0)[t]{${c^3}_2$}}
\put(160,0){\makebox(0,0)[t]{${c^4}_2$}}
\put(160,47){\makebox(0,0)[t]{$a$}}
\put(190,0){\makebox(0,0)[t]{${c^5}_2$}}
\put(190,50){\makebox(0,0)[t]{$b$}}
\put(220,0){\makebox(0,0)[t]{${c^2}_2$}}
\put(220,50){\makebox(0,0)[t]{$d$}}

\color{green}
\put(10,35){\circle*{5}}
\put(10,5){\circle*{5}}
\color{black}
\put(40,5){\circle*{5}}
\put(70,5){\circle*{5}}
\put(100,5){\circle*{5}}
\put(40,35){\circle*{5}}
\put(70,35){\circle*{5}}
\put(100,35){\circle*{5}}
\put(130,5){\circle*{5}}
\put(160,5){\circle*{5}}
\put(190,5){\circle*{5}}
\put(220,5){\circle*{5}}
\put(160,35){\circle*{5}}
\put(190,35){\circle*{5}}
\put(220,35){\circle*{5}}
\end{picture}
\end{center}
\caption{\label{PathClauseGadget} The clause gadget for the clause
  $({\sf a}
  \lor {\sf b} \lor {\sf d})$. Only at most two vertices from each empire are
  shown. In particular vertices labelled $T_1$ and $T_2$ are in
  $Z({\bf T})$,
  while vertices labelled $a$, $b$ and $d$ are in $Z({\bf a})$,
  $Z({\bf b})$ and
  $Z({\bf d})$ respectively.}
\end{figure}

Note that $P(\phi)$ is
$(3,r)$-colourable if and only if $\phi$ is satisfiable. This follows
from the properties of the well known 
reduction
3-SAT $\leq_p$ 3-COL, as 
the graph obtained from $P(\phi)$ by shrinking each connectivity
gadget first and then each remaining empire in $P(\phi)$ to
a distinct (pseudo-)vertex
 coincides with that created from $\phi$ using the
classical 3-COL reduction.
\qed}

 \bigskip

For
\margine{Case $s>3$. Starting point is WG's proof. However the
  connectivity gadgets have changed. So the proof will have some
  differences (esp. re: variable and clause gadgets).}
 $s>3$ the NP-hardness of  $s$-COL$_r$(LFOREST) follows from that of
 $s$-COL(FG$(s,s-1)$). The argument is much simpler than in the case
 described above. Given an $(s, s-1)$-formula graph $\Phi$, the
 $r$-empire graph obtained by linearizing  all vertices of $\Phi$ is
 an instance of $s$-COL$_r$(LFOREST). This immediately gives the
 following result.

\begin{ThS}
\label{Paths2}
Let $r$ and $s$ be fixed positive integers with $r \geq 3$, and 
$3 < s < 2r - \sqrt{2r + \frac{1}{4}}+ \frac{3}{2}$. Then
{\rm  $s$-COL(FG$(s,s-1)$) $\leq_p s$-COL$_r$(LFOREST)}.
\end{ThS}

 
\section{Trees}
\label{trees}

The result on linear forests of Section \ref{paths} already proves that 
$s$-COL$_r$ is NP-hard on planar graphs if $s \geq 3$ is sufficiently small.
In this section we investigate the effect of connectedness on
 the computational complexity of the
$s$-COL$_r$ problem.
The outcome of our investigation is the
following dichotomy result (in the next theorem TREE is the class of
all trees).

\begin{ThS}
\label{t1:trees}
 Let $r$ and $s$ be fixed positive integers with $r \geq 2$, then the
 {\rm $s$-COL$_r$(TREE)} problem
 is {\rm NP}-hard if $2 < s < 2r$, and polynomial time solvable otherwise.
\end{ThS}

The proof of Theorem \ref{t1:trees} is split into two parts.
The argument for $s=3$ is very
similar to the one we used for forests of paths, but simpler, as trees
are allowed to have vertices of arbitrary large degree. We
present the proof in some details only for the case $r=2$ (see Theorem \ref{r2s3-tree}
below). For $r>2$ note that a tree $T_1$ with empires 
of size $r_1$ can be translated into a tree $T_2$ with empires of size $r_2 > r_1$ by simply
attaching $r_2-r_1$ new leaves to a fixed element in each empire of
$T_1$. 
For $s>3$ we argue as in Section
 \ref{paths}, translating formula graphs into pairs formed by a tree
 and a partition of its vertices into empires. 
The hardness of $s$-COL$_r$(TREE)
 follows from Theorem \ref{fg}. Details in Theorem \ref{NPhard} below.

\begin{ThS}
\label{r2s3-tree}
{\rm 3-SAT $\leq_p$ 3-COL$_2$(TREE)}. 
\end{ThS}
\proof{(Sketch) Given an instance $\phi$ of 3-SAT we define
a tree $T(\phi)$ and a partition of its vertices into empires such that
$T(\phi)$ admits a
(3,2)-colouring if and only if $\phi$ is satisfiable. 
$T(\phi)$ will consist of one {\sl truth gadget}, one {\sl variable gadget} for each variable used in $\phi$, and one {\sl clause gadget} for each clause in $\phi$.

The truth gadget is a copy of $B^+_{2,2}({\bf T})$.
Since empires $\bf T$, $\bf F$ and $\bf X$ are adjacent to each other (in the gadget's reduced graph)
w.l.o.g. we assume they are coloured TRUE, FALSE and OTHER respectively.
 For each variable $\sf a$ in $\phi$, $T(\phi)$ contains a copy of
 $B_{2,2}$ spanned by empires labelled ${\bf a}$, ${\bf {\overline
     a}}$, and ${\bf X}$. The
construction
forces empires  ${\bf a}$, ${\bf {\overline
     a}}$ to be coloured differently from ${\bf X}$ (and each other).
Finally, for each clause in $\phi$, we use a clause gadget like the
one in Figure \ref{PathClauseGadget}.

Arguing like in the proof of Theorem \ref{r2s3} it is easy to see that
$T(\phi)$ is
$(3,2)$-colourable if and only if there is some way to assign the variables of
$\phi$ as TRUE or FALSE so that every clause contains at least
one TRUE literal.
\qed}

\medskip

\mz{2011.7.16}{In the next result $r \geq 3$????? So we do not have hardness for $r=2$
  and $s=4$?

IF THIS IS TRUE ALL ABSTRACTS MUST  BE CHANGED (WG + THIS ONE).

\medskip

IT'S FINE!! $r=2$ does not make sense if $s > 3$ as in that case $s
\geq 2r$ and the problem is easy!
}

\begin{ThS}
\label{NPhard}
{\rm $s$-COL(FG$(s,s-1))\leq_p s$-COL$_r$(TREE)}, for any $r \geq 3$ and $3 < s< 2r$. 
\end{ThS}
\proof{As in the proof of Theorem \ref{Paths2} we give a set of
  replacement rules that translate an $(s,s-1)$-formula graph $\Phi$
  into a tree $T(\Phi)$
 and a partition of $V(T(\Phi))$ into empires of size
  $r$ such that $T(\Phi)$
is $(s,r)$-colourable if and only if the formula graph is
$s$-colourable. This time there is no need to use the connectivity 
gadgets $A_{r,s,m}$ as the vertices of $T(\Phi)$ can have arbitrarily
large degrees. However some care is needed to make sure that the
resulting graph is in fact a tree. 

In details, the complete graph on $\{T,F,X^1, \ldots, X^{s-2}\}$ is
replaced by a copy of $B^+_{r,s-1}({\bf T})$ with empires labelled
${\bf T}$, ${\bf F}$,
and ${\bf X}^1, \ldots, {\bf X}^{s-2}$. Note that, as discussed in
Section \ref{G}, this graph is in
fact a tree (Figure \ref{FrsPlusMinus} displays the connected clique gadget for
$r=3$ and $s=5$). 
Also, because of constraint {\bf B3}
in the definition of $B_{r,s}$, w.l.o.g. 
we may assume that colours ``TRUE", ``FALSE", ``OTHER$^1$", $\ldots$,
``OTHER$^{s-2}$" are assigned to empires ${\bf T}$, ${\bf F}$, ${\bf X}^1
\ldots, {\bf X}^{s-2}$
respectively. 
\begin{figure}[t]
\begin{center}
\includegraphics[scale=0.5]{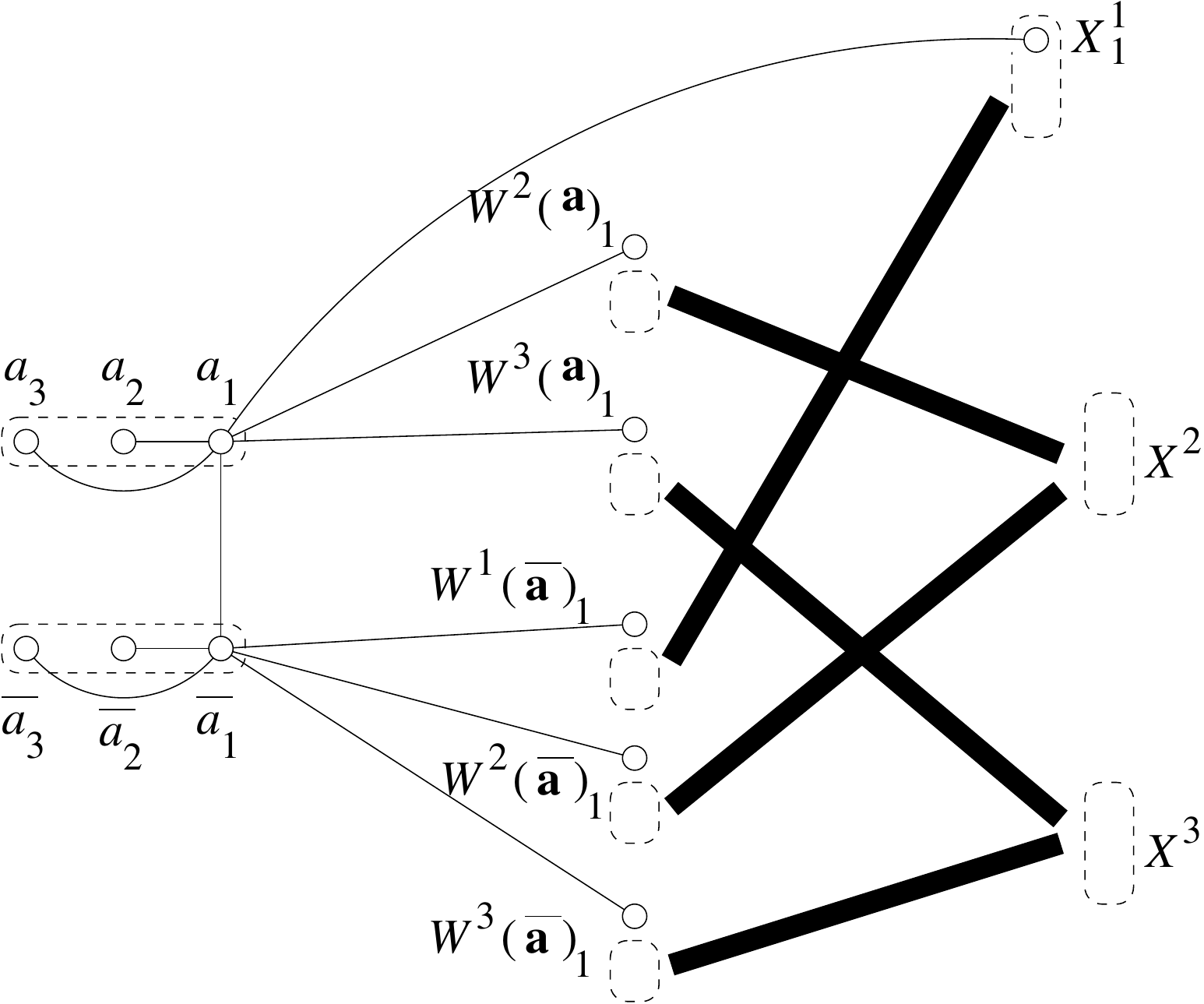}
\vspace{-.3cm}
\caption{\label{Variable} The gadget for the complementary pair $a$ and
  $\overline a$ when $r=3$, $s=5$. The dashed blobs represent either
  empires, or part of them. The diagram clearly shows all copies of 
$B^-_{3,5}({\bf W}^i(\ell),{\bf X}^i)$, following the graphical
notation introduced in Figure \ref{ccg}.}
\end{center}
\end{figure}

 For each complementary pair $a, \overline{a}$ of $V(\Phi)$ we 
create $2s-5$ empires ${\bf W}^2(a), \ldots, {\bf W}^{s-2}(a)$ and
${\bf W}^1(\overline{a}), \ldots, {\bf W}^{s-2}(\overline{a})$. These are then
connected to $B^+_{r,s-1}({\bf T})$ using the graphs
$B^-_{r,s}({\bf W}^i(a),
{\bf X}^i)$, and
$B^-_{r,s}({\bf W}^i(\overline{a}), {\bf X}^i)$ for all 
$i \in  \{1, \ldots, s-2\}$. 
For each $a \in {\cal A}$ the subgraph of $\Phi$ spanned by
$\bigcup_i \{a,\overline{a},X^i\}$ is represented by a graph like
the one sketched in Figure \ref{Variable} for $r=3$ and $s=5$. This graph involves empires $\bf a$,
${\bf \overline{a}}$, ${\bf X}^1, \ldots, {\bf X}^{s-2}$, ${\bf W}^2(a), \ldots, {\bf W}^{s-2}(a)$ and
${\bf W}^1(\overline{a}), \ldots, {\bf W}^{s-2}(\overline{a})$. 
Empires $\bf a$, and 
${\bf \overline{a}}$, each span a tree with one vertex, w.l.o.g. $a_1$
(resp. $\overline{a}_1$) of degree $r-1$ and $r-1$ vertices of degree
one, all adjacent to it. These two trees are connected by the edge
$\{a_1,\overline{a}_1\}$. Vertex $a_1$ (resp. $\overline{a}_1$) is
also connected to the vertex in ${\bf W}^2(a), \ldots, {\bf
  W}^{s-2}(a)$ left isolated in the graph $B^-_{r,s}({\bf W}^i(a),
{\bf X}^i)$ (resp. to the isolated vertex in ${\bf W}^1(\overline{a})_1, \ldots, {\bf
  W}^{s-2}(\overline{a})_1$ belonging to $B^-_{r,s}({\bf W}^i(\overline{a}),
{\bf X}^i)$). Finally $a_1$ is  connected to
$X^1_1$.
The edge $\{a_1,X^1_1\}$ ensures that the union of $B^+_{r,s-1}({\bf
  T})$ and 
the graph spanned by empires 
 $\bf a$,
${\bf \overline{a}}$, ${\bf X}^1$, ${\bf W}^2(a), \ldots, {\bf W}^{s-2}(a)$ and
${\bf W}^1(\overline{a}), \ldots, {\bf W}^{s-2}(\overline{a})$
is just a single tree. The edges connecting empires  $\bf a$,
${\bf \overline{a}}$, with ${\bf W}^2(a), \ldots, {\bf W}^{s-2}(a)$ and
${\bf W}^1(\overline{a}), \ldots, {\bf W}^{s-2}(\overline{a})$,
because of the properties of the $({\bf W}^i(\ell),{\bf X}^i)$-colour constraining gadgets, 
prevent   ${\bf a}$ and ${\bf {\overline
    a}}$ from being able to use the colours of the ${\bf X}^i$ in any
colouring of $T(\Phi)$.

Each group $\{c^{1}, \ldots, c^{s-1}\}$ in ${\cal C}$ is replaced by
empires ${\bf c}^{1}, \ldots, {\bf
  c}^{s-1}$ (different groups replaced by different sets of empires).  
The complete graph on
$\{T,c^{1}, \ldots, c^{s-1}\}$ 
is replaced by a copy of 
$B_{r,s-1}$ on the corresponding empires (this ensures that the union
of 
$B^+_{r,s-1}({\bf T})$ and such $B_{r,s-1}$ form a single tree).
We then attach to this graph $s-1$ graphs
$B^-_{r,s}({\bf b}^{j},{\bf c}^{j})$, for $j \in \{1, \ldots,
s-1\}$. Empire ${\bf b}^{j}$ must have the same colour as ${\bf
  c}^{j}$ and it has, in $B^-_{r,s}({\bf b}^{j},{\bf c}^{j})$,
an isolated vertex, $b^{j}_1$. If $\ell$ is the unique element of $\cal A$
adjacent to $c^{j}$ in the formula graph then $\{b^{j}_1, \ell_1\}$ is  an edge of $T(\Phi)$.
A schematic representation of the subgraph induced by {\bf T}, 
empires
${\bf c}^1, \ldots, {\bf c}^{s-1}$, along with the copies of
$B^-_{r,s}({\bf b}^{j},{\bf c}^{j})$
is given in Figure \ref{cl}.
\begin{figure}[htb]
\begin{center}
\includegraphics[scale=0.5]{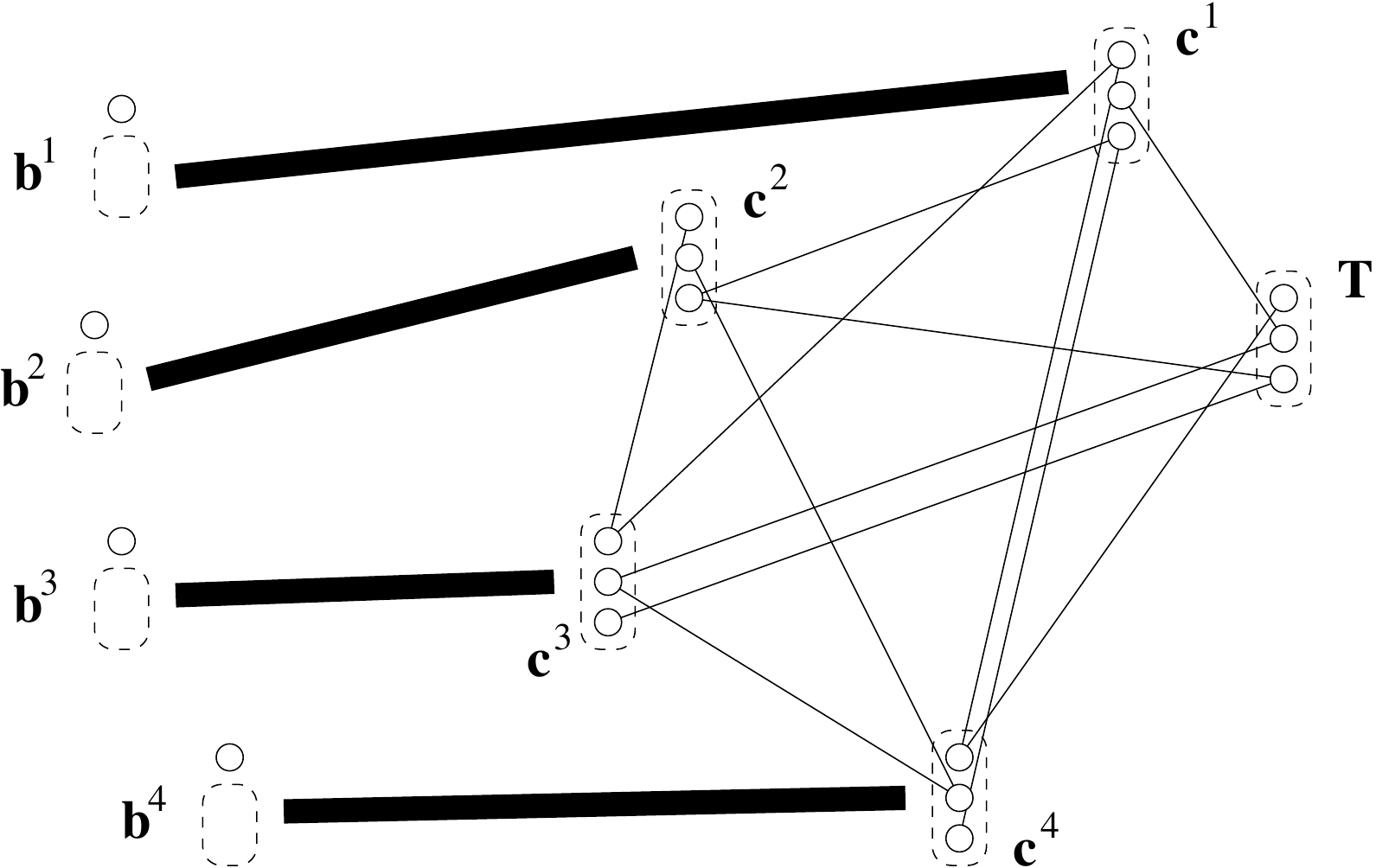}
\end{center}
\caption{\label{cl} A schematic representation of {\bf T}, 
empires
${\bf c}^1, \ldots, {\bf c}^{s-1}$, all edges among these along with the copies of
$B^-_{r,s}({\bf b}^{j},{\bf c}^{j})$.}
\end{figure}

The overall construction is such that for each vertex in $V(\Phi)$
there is an equivalent empire in $V(T(\Phi))$, and for each edge in
$E(\Phi)$ there is an edge $\{u,v\} \in E(T(\Phi))$ that either
connects the corresponding empires $\bf u$ and $\bf v$ or connects
$\bf u$ to an empire that must be given the same colour as $\bf v$ in
any $(s,r)$-colouring
of $T(\Phi)$. From this we can see that $T(\Phi)$ admits an $(s,r)$-colouring if and only if $\Phi$ admits an $s$-colouring.
\qed}

\section{General Planar Graphs}
\label{planar}

Theorem \ref{t1:trees} of last section does not exclude the possibility
that $s$-COL$_r$ be solvable in polynomial time for arbitrary planar
graphs provided $s \geq 2r$. Here we show that in fact this
is not the case.
The main result of this section is the following:

\begin{ThS}
\label{t2:planar}
 Let $r$ and $s$ be fixed positive integers with $r \geq 2$, then the
 {\rm $s$-COL$_r$} problem is {\rm NP}-hard if $3 \leq s <
 6r-3-2\delta_{r,2}$, and solvable in polynomial time if $s=2$ or $s
 \geq 6r$.
\end{ThS}

Note that $s$-COL$_r$ can be solved in polynomial time for $s=2$
(as checking if the reduced graph of a planar graph 
is bipartite is easy) and for $s \geq 6r$ (because
of Heawood's result). Also, Theorem \ref{t1:trees} proves the case $s
< 2r$. Therefore only the case $s \geq 2r$ needs further discussion.
The bulk of the argument
is similar to that of Theorem \ref{Paths2} and \ref{NPhard} with a
couple of differences. First,  this time we only need the graph
resulting from the transformation of the initial formula graph to be
planar (note that the formula graph in general is NOT planar). On the
other hand,
we want the transformation to work for much larger values of $s$. 
Our solution hinges on proving that all complete subgraphs of the
starting formula graph and a number of other gadgets
attached to them have sufficiently large thickness. For the complete
graphs we may use well-known results \cite{beineke97:_biplan}, whereas
for 
the specific
gadgets we need a bespoke construction. 

Using the gadgets descrived above we can prove the following result, which
completes the proof of Theorem \ref{t2:planar}. 

\begin{ThS}
\label{NPhardPlanar}
{\rm $s$-COL(FG$(s,s-1))\leq_p s$-COL$_r$}, for any $r \geq 2$ and $2r
\leq s < 6r-3- 2\delta_{r,2}$.
\end{ThS}
\proof{The proof mirrors that of Theorem \ref{NPhard}.
We once again give a set of replacement rules to convert a $(s,s-1)$-formula graph $\Phi$ into a planar graph $G(\Phi)$ that is $(s,r)$-colourable if and only if $\Phi$ is $s$-colourable.

The copy of $K_s$ induced by the vertex set $\cal T$ in $\Phi$ 
is replaced by $r$ edge disjoint subgraphs of $K_s$.
For $s \leq 6r-4$ 
the existence of such graphs is granted by known results on the thickness of $K_s$
\cite{beineke97:_biplan}. W.l.o.g. we may assume that the empires of
the resulting graph (which, as usual, we label ${\bf T}$, ${\bf F}$,
and ${\bf X}^1,
{\bf X}^2, \ldots$) are coloured ``TRUE", ``FALSE", ``OTHER$^1$", $\ldots$,
``OTHER$^{s-2}$" respectively. The graph is then expanded, for each
$a, \overline{a} \in {\cal A}$ using empires ${\bf W}^2(a),
\ldots,{\bf  W}^{s-2}(a)$ and
${\bf W}^1(\overline{a}), \ldots, {\bf W}^{s-2}(\overline{a})$ and the graphs
$D_{r,s}({\bf W}^i(a), {\bf X}^i)$  for all $i$ such that $2 \leq i
\leq s-2$, and $D_{r,s}({\bf W}^i(\overline{a}), {\bf X}^i)$ for all $i$ such that
$1 \leq i \leq s-2$.
The graphs $\Phi[\bigcup \{a,\overline{a},X^j\}]$ and 
$\Phi[\{T\} \cup {\cal C}]$ are subject to transformations similar to
those in Theorem \ref{NPhard} but using the planar decomposition of
the complete graph instead of copies of $B_{r,s}$ and graphs
$D_{r,s}({\bf u},{\bf v})$
instead of $B_{r,s}^-({\bf u}, {\bf v})$.

As in Theorem \ref{NPhard}, every vertex in
$V(\Phi)$ has a corresponding empire in $V(G(\Phi))$, and every edge
$\{u, v\} \in E(\Phi)$ has a corresponding edge in $E(G(\Phi))$ that
connects either the empires ${\bf u}$ and ${\bf v}$ or empires that must be given the same colour as them in any proper $(s,r)$-colouring. It follows that $G(\Phi)$ admits a proper $(s,r)$-colouring if and only if $\Phi$ admits a proper $s$-colouring.
\qed}

The reduction in the proof of Theorem \ref{NPhardPlanar} shows that
for any given formula graph $\Phi$ one can define a planar graph
$G(\Phi)$ which is formed by (at least) $r$ connected components and
reduces to $\Phi$. Thus the proof is actually showing, for $s
\geq 2r$, the
NP-hardness of colouring, in the traditional sense, graphs of
thickness $r$. The following result can be obtained extending the
proof to any $s>3$ and using a more direct reduction from 3-SAT for $s=3$.

\begin{ThS}
It is {\rm NP}-hard to decide
whether a graph of thickness $r>1$ can be coloured with $s < 6r - 3 - 2\delta_{r,2}$ colours. 
\end{ThS}

An obvious way to improve Theorem \ref{NPhardPlanar} (and perhaps close the small gap
between NP-hard and polynomially decidable cases) would be to use
different gadgets to replace the complete subgraphs of $\Phi$. However,
it seems difficult to devise a graph with high thickness that shares the
colour constraining properties of the complete graph.
Perhaps, a more direct reduction from
the satisfiability problem may provide a handle on the remaining open cases.

\bibliography{../../../../../lib/bib/references.bib}

\begin{thebibliography}{10}


\bibitem{beineke97:_biplan}
L.~W. Beineke.
\newblock Biplanar graphs: a survey.
\newblock {\em Computers and Mathematical Applications}, 34(11):1--8, 1997.


\bibitem{beineke65}
L.~W. Beineke and F.~Harary.
\newblock The thickness of the complete graph.
\newblock {\em Canadian Journal of Mathematics}, 17:850--859, 1965.

\bibitem{bhm64}
L.~W. Beineke, F.~Harary, and J.~W. Moon.
\newblock On the thickness of the complete bipartite graph.
\newblock {\em Proceedings of the Cambridge Philosophical Society}, 60(1):1--5,
  1964.

\bibitem{B41}
R.~L. Brooks.
\newblock On colouring the nodes of a network.
\newblock {\em Proc. Cambridge Phil. Soc.}, 37:194--197, 1941.

\bibitem{bryant07:_cycle}
D.~E. Bryant.
\newblock Cycle decompositions of the complete graphs.
\newblock In A.~J.~W. Hilton and J.~M. Talbot, editors, {\em Surveys in
  Combinatorics}, volume 346 of {\em London Mathematical Society Lecture Notes
  Series}, pages 67--97. Cambridge University Press, 2007.

\bibitem{cooper09:_martin_trees_and_empir_chrom}
C.~Cooper, A.~R.~A. McGrae, and M.~Zito.
\newblock Martingales on trees and the empire chromatic number of random trees.
\newblock In M.~Kuty\mbox{\l}owski, M.~Gebala, and W.~Charatonik, editors, {\em
  FCT 2009}, volume 5699 of {\em Lecture Notes in Computer Science}, pages
  74--83. Springer-Verlag, 2009.

\bibitem{cormen09:_introd_algor}
T.~H. Cormen, C.~E. Leiserson, R.~L. Rivest, and C.~Stein.
\newblock {\em Introduction to Algorithms}.
\newblock M.I.T. Press, third edition, 2009.

\bibitem{diestel99:_graph_theor}
R.~Diestel.
\newblock {\em Graph Theory}, volume 173 of {\em Graduate Texts in
  Mathematics}.
\newblock Springer-Verlag, 1999.

\bibitem{GJ79}
M.~R. Garey and D.~S. Johnson.
\newblock {\em Computer and Intractability, a Guide to the Theory of
  {NP-C}ompleteness}.
\newblock Freeman and Company, 1979.

\bibitem{G}
A.M. Gibbons.
\newblock {\em Algorithmic Graph Theory}.
\newblock Cambridge University Press, 1985.

\bibitem{heawood90:_map}
P.~J. Heawood.
\newblock Map colour theorem.
\newblock {\em Quarterly Journal of Pure and Applied Mathematics}, 24:332--338,
  1890.

\bibitem{hutchinson93:_color}
J.~P. Hutchinson.
\newblock Coloring ordinary maps, maps of empires, and maps of the moon.
\newblock {\em Mathematics Magazine}, 66(4):211--226, 1993.

\bibitem{jackson84:_solut}
B.~Jackson and G.~Ringel.
\newblock Solution of {H}eawood's empire problem in the plane.
\newblock {\em Journal f\"ur die Reine und Angewandte Mathematik},
  347:146--153, 1983.

\bibitem{K72}
R.~M. Karp.
\newblock Reducibility among combinatorial problems.
\newblock In Raymond~E. Miller and James~W. Thatcher, editors, {\em Complexity
  of Computer Computations (Proceedings of a Symposium on the Complexity of
  Computer Computations, March, 1972, Yorktown Heights, NY)}, pages 85--103.
  Plenum Press, New York, 1972.

\bibitem{lovasz93:_combin_probl_exerc}
L.~Lov\'asz.
\newblock {\em Combinatorial Problems and Exercises}.
\newblock North-Holland, second edition, 1993.

\bibitem{lucas92:_recreat_mathem_vol}
E.~Lucas.
\newblock {\em R\'ecreations Math\'ematiqu\'es (Vol. II)}.
\newblock Gauthier-Villars, 1892.

\bibitem{maekinen09}
Erkki M\"akinen and T.~Poranen.
\newblock {\em An annotated bibliography on the thickness, outerthickness, and
  arboricity of a graph}, volume D-2009-3 of {\em D-NET PUBLICATIONS}.
\newblock University of Tampere, 2009.

\bibitem{mcgrae08:_colour_random_empir_trees}
A.~R. McGrae and M.~Zito.
\newblock Colouring random empire trees.
\newblock In E.~Ochma\'nski and J.~Tyszkiewicz, editors, {\em Mathematical
  Foundations of Computer Science 2008}, volume 5162 of {\em Lecture Notes in
  Computer Science}, pages 515--526. Springer-Verlag, 2008.

\bibitem{mcgrae10:_colour_empir_random_trees}
A.~R.~A. McGrae.
\newblock {\em Colouring Empires in Random Trees}.
\newblock PhD thesis, Department of Computer Science, University of Liverpool,
  2010.
\newblock Available from {\footnotesize {\tt
  http://www.csc.liv.ac.uk/research/techreports/techreports.html}} as technical
  report ULCS-10-007.

\bibitem{mutzel98}
P.~Mutzel, T.~Odenthal, and M.~Scharbrodt.
\newblock The thickness of graphs: a survey.
\newblock {\em Graphs and Combinatorics}, 14:59--73, 1998.

\end{thebibliography}
\end{document}